\documentclass[aps,reprint,superscriptaddress,prb,longbibliography]{revtex4-2}
\usepackage{amsmath}
\usepackage{amssymb}
\usepackage{graphicx}
\usepackage[english]{babel}
\usepackage{bm}
\usepackage{color}
\usepackage{hyperref}
\hypersetup{
    colorlinks=true,
    linkcolor=blue,
    citecolor=blue,      
    urlcolor=blue,
}

\newcommand{\epsvac}{\varepsilon_0} % Vacuum permittivity, \varepsilon_{\text{vac}} is also possible
\newcommand{\dee}{\mathrm{d}} % Integration variable, such as \dee x

\newcommand{\eref}[1]{Eq.~(\ref{#1})}  % Reference to one equation
\newcommand{\erefs}[2]{Eqs.~(\ref{#1})-(\ref{#2})}  % Reference to multiple equations
\newcommand{\figref}[1]{Fig.~\ref{#1}}  % Reference to one figure
\newcommand{\figrefs}[2]{Figs.~\ref{#1}-\ref{#2}}  % Reference to multiple figures
\newcommand{\secref}[1]{Sec.~\ref{#1}}  % Reference to one section
\newcommand{\citeref}[1]{Ref.~\cite{#1}}  % Cite one paper
\usepackage[normalem]{ulem} 

\usepackage{xr}  % cross-referencing between two documents
\definecolor{changecolor}{rgb}{0.3,0.3,1.0}%
\begin{document}

\title{
Importance of nonlinear long-range electron-phonon interaction for the carrier mobility of anharmonic halide perovskites
}

\author{Matthew Houtput}
\affiliation{Theory of Quantum Systems and Complex Systems, Universiteit Antwerpen, B-2000 Antwerpen, Belgium}

\author{Ingvar Zappacosta}
\affiliation{Theory of Quantum Systems and Complex Systems, Universiteit Antwerpen, B-2000 Antwerpen, Belgium}

\author{William Wenborn}
\affiliation{Theory of Quantum Systems and Complex Systems, Universiteit Antwerpen, B-2000 Antwerpen, Belgium}

\author{Serghei Klimin}
\affiliation{Theory of Quantum Systems and Complex Systems, Universiteit Antwerpen, B-2000 Antwerpen, Belgium}

\author{Samuel Ponc\'e}
\affiliation{European Theoretical Spectroscopy Facility and Institute of Condensed Matter and Nanosciences, Universit\'e catholique de Louvain, Chemin des \'Etoiles 8, B-1348 Louvain-la-Neuve, Belgium}
\affiliation{WEL Research Institute, avenue Pasteur 6, 1300 Wavre, Belgium}

\author{Jacques Tempere}
\affiliation{Theory of Quantum Systems and Complex Systems, Universiteit Antwerpen, B-2000 Antwerpen, Belgium}

\author{Cesare Franchini}
\affiliation{Faculty of Physics, Computational Materials Physics, University of Vienna, Kolingasse 14-16, Vienna A-1090, Austria}
\affiliation{Department of Physics and Astronomy, Alma Mater Studiorum - Universit\`a di Bologna, Bologna, 40127 Italy}

\begin{abstract}
The interaction between the electrons and the lattice vibrations in a solid is responsible for various important effects, such as formation of polarons, temperature dependent bandgaps, phonon-limited carrier transport, and conventional superconductivity. 
Most works assume a linear electron-phonon interaction, where the electron only interacts with one phonon at a time. 
However, the validity of this assumption has not been verified in polar anharmonic materials, where large ionic displacements may invalidate the assumption of linear interaction. 
Here, we show that nonlinear electron-phonon interactions contribute significantly to the finite-temperature electron mobility of the inorganic lead halide perovskite CsPbI$_3$. 
The effect of nonlinear interaction is taken into account using the recently derived expression for the long-range part of the one-electron-two-phonon matrix element.
We calculate the electron mobility from first principles within the self-energy relaxation-time approximation, treating the electron–phonon coupling in the long-range approximation.
Despite these approximations, the calculated mobilities are in good agreement with the available experimental data, while enabling us to isolate and quantify the contribution of nonlinear electron–phonon interactions relative to the conventional linear coupling.
We find that the one-electron–two-phonon interaction modifies the temperature dependence of the mobility in CsPbI$_3$ and reduces its room-temperature value by about 10\%.
This sizeable contribution results from the combined effects of strong lattice anharmonicity and large thermal phonon populations, the latter being enhanced by the low phonon frequencies associated with the heavy constituent atoms.
These results identify a regime in which nonlinear electron–phonon interactions become relevant and indicate that they should be explicitly considered for finite-temperature properties of anharmonic halide perovskites.
\end{abstract}

\maketitle

Halide perovskites \cite{ranalli2024}  are known for their excellent optoelectronic properties~\cite{jang2015, stranks2015, brenner2016}. 
This has led to their usage in the development of several promising technological applications, such as solar cells~\cite{kojima2009, green2014, min2021} and energy-efficient lighting~\cite{tan2014, liu2021}. 
A particularly important property in the context of these applications is the charge transport~\cite{ponce2019}, usually quantified by the carrier mobility $\mu$. 
A deeper understanding of the microscopic processes that govern mobility is therefore desirable.

\begin{figure}
\centering
\includegraphics[width=8.6cm]{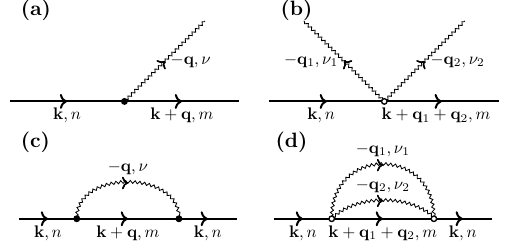}
\caption{\label{fig:Diagrams} \textbf{(a)}-\textbf{(b)} Feynman diagrams for the linear and lowest-order nonlinear electron-phonon interaction. 
In \textbf{(a)}, the long-range approximation amounts to $\mathbf{q}\rightarrow \mathbf{0}$, and in \textbf{(b)} it amounts to $-\mathbf{q}_1 \rightarrow \mathbf{q}_2$. 
\textbf{(c)}-\textbf{(d)} Contributions to the imaginary part of the retarded self energy that are taken into account in this article, where \textbf{(c)} is the Fan-Migdal diagram, and \textbf{(d)} is its nonlinear analog.
We do not include the Debye-Waller diagram since it is real.
}
\end{figure}

In most materials, this microscopic understanding can be found from the theory of phonon-limited transport \cite{ponce2020}. 
One assumes that the charge carriers can interact with harmonic phonons with frequencies $\omega_{\mathbf{q}\nu}$, through a linear electron-phonon scattering process as shown in \figref{fig:Diagrams}a. 
Then, a common way to calculate the carrier mobility $\mu$ is to calculate the electron energies $\varepsilon_{\mathbf{k}n}$, phonon frequencies $\omega_{\mathbf{q}\nu}$, and electron-phonon matrix elements $g_{mn\nu}(\mathbf{k},\mathbf{q})$ from first principles \cite{giustino2017}, and use these to solve the linearized Boltzmann transport equation for the carrier mobility \cite{ponce2020}. 
However, many of the assumptions in this theory are not applicable to most halide perovskites. 
For example, it has been shown that several of the halide perovskites are strongly anharmonic \cite{yaffe2017, gold-parker2018, schilcher2021, zhu2025, yin2025}, requiring a treatment with renormalized phonon frequencies and spectral broadening, or even the full phonon spectral function. 
It has also been suggested that the quasiparticle picture might not be valid in halide perovskites \cite{schilcher2021}, requiring a treatment beyond the Boltzmann transport equation \cite{lihm2026}. 
The violation of these common assumptions makes comparison with experimental measurements of the mobility \cite{yi2016, dastidar2017, shrestha2018, hutter2018, biewald2019, sarkar2021, xia2021, zhang2021, lee2022} difficult.

We show here that another common assumption is violated in halide perovskites: the assumption of linear electron-phonon interaction. 
This approximation is often made out of necessity, as the electron-phonon matrix elements for higher-order electron-phonon interactions such as the one in \figref{fig:Diagrams}b are not known. 
Regardless, it has been predicted that some halide perovskites get significant contributions to the band gap renormalization from nonlinear electron-phonon interaction~\cite{saidi2016, mayers2018}. 

In this Letter, we quantify the effect of the nonlinear one-electron-two-phonon interaction on the carrier mobility of cesium lead iodide (CsPbI$_3$), a representative inorganic halide perovskite. 
There have been several treatments of nonlinear electron-phonon interaction in the literature in recent years \cite{ngai1974, epifanov1981, vandermarel2019, kiselov2021, kumar2021, nazaryan2021, bianco2023, yildirim2001, liu2001}, but most focus on lattice model Hamiltonians \cite{riseborough1984, adolphs2013, adolphs2014, li2015, li2015a, kennes2017, sentef2017, dee2020, grandi2021, sous2021, prokofev2022, ragni2023, zhang2023, han2024, kovac2024, klimin2024, ragni2025}. We note that nonlinear interaction should not be confused with higher-order treatments of the linear interaction, such as those which include the crossing diagrams \cite{smondyrev1986, lee2020}.
In order to treat the nonlinear interaction from first principles, we use the recently derived expression for the dipole long-range part of the one-electron-two-phonon matrix element $g_{mn \nu_1 \nu_2}(\mathbf{k},\mathbf{q}_1,\mathbf{q}_2)$ ~\cite{houtput2025}. 
All material properties that enter this matrix element can be calculated from first principles. 
Since the linear electron-phonon interaction in CsPbI$_3$ is dominated by the long-range contribution~\cite{ponce2019, xia2021, yamada2022}, we expect this analysis to be representative for the full nonlinear one-electron-two-phonon interaction in CsPbI$_3$. We investigate the accuracy of the long-range approximation further in \secref{app:LRbenchmarking} of the Supplemental material \cite{SIref}.

Our aim is to establish physical criteria for determining when nonlinear electron–phonon interactions can be safely neglected and when they must be taken into account.
Such criteria provide a useful framework for assessing previous and future calculations of halide perovskites that include only linear electron–phonon coupling \cite{ponce2019, lafuente-bartolome2024, yin2025, zhu2025}.
To this end, we investigate the relative contributions of the linear and nonlinear electron-phonon interaction to the carrier mobility in the cubic phase of CsPbI$_3$.
In order to use the theory derived in \citeref{houtput2025}, we use the harmonic and quasiparticle approximations.
As discussed, these are likely poor approximations in CsPbI$_3$.
Furthermore, we calculate the mobility using the self-energy relaxation time approximation (SERTA), rather than solving the linearized Boltzmann Transport Equation (BTE) iteratively.
While these approximations may affect the absolute values of the calculated mobilities, they provide a consistent framework for assessing the relative importance of linear and nonlinear electron–phonon scattering.
Reassuringly, the resulting mobilities are also in good agreement with the available experimental data.
Potential directions for a relaxation of these assumptions are discussed at the end of the manuscript.

The drift electron mobility $\mu_e$ can be computed with the linearized BTE~\cite{ponce2020}:
\begin{equation} \label{muSERTA}
\mu_{e,\alpha \beta} = \frac{-e}{n_e (2\pi)^3} \!\!\sum_{n \in \text{CB}}  \int_{\text{1BZ}} \!\!\!\! \dee^3\mathbf{k} v_{\mathbf{k}n\alpha}  \partial_{E_\beta}f(\varepsilon_{\mathbf{k}n}). 
\end{equation}
Here $\alpha, \beta \in \{x,y,z\}$ represent Cartesian directions, $e$ is the elementary charge, $n_e$ is the electron density, $\hbar \mathbf{k}$ is the crystal momentum and is restricted to the first Brillouin zone (1BZ), $n$ labels the electron bands restricted to the conduction bands, $\varepsilon_{\mathbf{k}n}$ are the electron band energies, and $v_{\mathbf{k}n\alpha} = \frac{1}{\hbar} \frac{\partial \varepsilon_{\mathbf{k}n}}{\partial k_{\alpha}}$ are the band velocities.
Finally, the most important quantity that appears in \eref{muSERTA} is the out-of-equilibrium occupation function $\partial_{E_\beta}f(\varepsilon_{\mathbf{k}n})$ due to an applied electric field $E_\beta$ applied in the direction $\beta$. 

In the SERTA~\cite{ponce2018a}, the out-of-equilibrium electronic occupation functions reduce to:
\begin{equation} \label{ooeoF}
   \partial_{E_\beta}f(\varepsilon_{\mathbf{k}n}) = \frac{\partial f^0(\varepsilon_{\mathbf{k}n})}{\partial \varepsilon_{\mathbf{k}n}}  v_{\mathbf{k}n\beta} \tau_{\mathbf{k}n}, 
\end{equation}
Here, $f^0(\varepsilon)$ is the equilibrium occupation function for the electrons, which is the Fermi-Dirac distribution. 
Additionally, $\tau_{\mathbf{k}n}^{-1}$ are the scattering rates of the electrons with the phonons~\cite{ponce2020}, which can be obtained from the on-shell retarded diagonal electron self energy $\Sigma^{\rm{R}}_{\mathbf{k}nn}$~\cite{ponce2020}:
\begin{equation} \label{tauDef}
\frac{1}{\tau_{\mathbf{k}n}} = - \frac{2}{\hbar} \text{Im}\left[ \Sigma^{\rm{R}}_{\mathbf{k}nn}(\varepsilon_{\mathbf{k}n})\right]
\end{equation}
If we were to only use the linear electron-phonon interaction, we would be able to use the standard expression for the Fan-Migdal self energy from \figref{fig:Diagrams}c, written in terms of $\varepsilon_{\mathbf{k}n}$, $\omega_{\mathbf{q}\nu}$, and $g_{mn\nu}(\mathbf{k},\mathbf{q})$ \cite{ponce2020}.
In that case, the BTE has also been derived \cite{ponce2020} and implemented \cite{lee2023}, but this is not the case for the Hamiltonian that contains a nonlinear electron-phonon matrix element $g_{mn\nu_1\nu_2}(\mathbf{k},\mathbf{q}_1,\mathbf{q}_2)$.
The advantage of using SERTA is that it only requires the self energy, so we avoid having to rederive and reimplement the BTE.
Instead, we can use the long-range theory from \citeref{houtput2025}, where the retarded self energy was calculated up to lowest order in the electron-phonon interaction.
This gives the Fan-Migdal self energy, \figref{fig:Diagrams}c, and a similar term due to the nonlinear electron-phonon interaction in \figref{fig:Diagrams}d \cite{houtput2025}:
\begin{multline}\label{SelfEnergy}
\Sigma^{\rm R}_{\mathbf{k}nn}(\varepsilon_{\mathbf{k}n}) = \frac{\hbar e^2}{\epsvac (2\pi)^3} \int_{0}^{+\infty} \!\!\!\! \dee \omega \int_{\mathbb{R}^3} \!\! \dee^3 \mathbf{Q} \frac{\mathbf{Q} \cdot \bm{\mathcal{P}}(\omega) \cdot \mathbf{Q} }{(\mathbf{Q}\cdot \bm{\varepsilon}^{\infty} \cdot \mathbf{Q})^2} \\
\times \bigg[ \frac{1 - f^0(\varepsilon_{\mathbf{k}+\mathbf{Q}n}) + n(\hbar\omega)}{\varepsilon_{\mathbf{k}n} - \varepsilon_{\mathbf{k}+\mathbf{Q}n} - \hbar \omega + i\delta} + \frac{f^0(\varepsilon_{\mathbf{k}+\mathbf{Q}n}) + n(\hbar\omega)}{\varepsilon_{\mathbf{k}n} - \varepsilon_{\mathbf{k}+\mathbf{Q}n} + \hbar \omega + i\delta} \bigg]. 
\end{multline}
In this expression, $n(\hbar\omega)$ is the Bose-Einstein distribution for phonons, and $\mathbf{Q} = \mathbf{q}+\mathbf{G}$ represents a phonon momentum extended to the entire reciprocal space $\mathbb{R}^3$: this is equivalent to an integral over $\mathbf{q} \in 1\text{BZ}$ and a sum over reciprocal lattice vectors $\mathbf{G}$.
We note that the integral over $\mathbf{Q}$ should decay quickly and approximately vanish at the boundary of the first Brillouin zone in order for the long-range approximation to be valid, so in practice it is irrelevant whether the integration domain is $\mathbb{R}^3$ or $1\text{BZ}$.
$\bm{\varepsilon}^{\infty}$ and $\bm{\mathcal{P}}(\omega)$ are material-specific 3$\times$3 tensors:
$\varepsilon^{\infty}_{ij}$ is the high-frequency dielectric tensor, and $\mathcal{P}_{ij}(\omega)$ is a dimensionless spectral function which represents the strength of the long-range electron-phonon interaction.
For both tensors, the Cartesian directions $i$ and $j$ refer to the direction of a constant electric field $\mathcal{E}_i$ or $\mathcal{E}_j$ which is induced by the conduction electrons.
\begin{figure*}[ht]
\centering
\includegraphics[width=16.2cm]{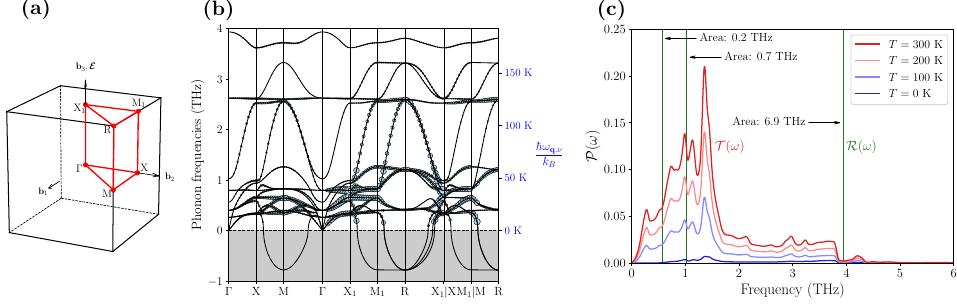}
\caption{\label{fig:Inputs} First principles results for the nonlinear electron-phonon properties of CsPbI$_3$ in the cubic phase. \textbf{(a)} High-symmetry Brillouin zone path for a simple cubic material, with an electric field breaking the symmetry in the $z$-direction. \textbf{(b)} Phonon frequencies $\omega_{\mathbf{q}\nu}$ of cubic CsPbI$_3$ at zero temperature. 
The overlaid circles have areas proportional to $\sum_{\nu'}|Y_{\nu\nu'z}(\mathbf{q})|^2$, i.e. the magnitude of the long-range interaction strengths $|Y_{\nu\nu'z}(\mathbf{q})|^2$ between the phonon $(\mathbf{q}\nu)$ and all other phonons of the form $(-\mathbf{q}\nu')$. There are imaginary phonon modes because the cubic phase is unstable at zero temperature: we set $Y_{\nu\nu',z}(\mathbf{q}) = 0$ for any branches with $\omega_{\mathbf{q},\nu} < 0.1~\text{THz}$. \textbf{(c)} One-electron-two-phonon spectral function $\mathcal{T}(\omega)$ calculated from \eref{TomegaDef} and the data from \textbf{(b)}. 
$\mathcal{R}(\omega)$ is shown as a sequence of vertical lines representing the Dirac delta functions, with the areas under the delta functions indicated for reference. The effect of the one-electron-two-phonon interaction is significantly enhanced at finite temperatures: this is ultimately due to the very low phonon frequencies of CsPbI$_3$ (see main text).
}
\end{figure*}
Using the theory from \citeref{houtput2025}, the electron-phonon spectral function can be calculated as a sum of two contributions.
The first contribution is $\mathcal{T}_{ij}(\omega)$, which is the contribution from the one-electron-two-phonon interaction. It originates from the diagram in \figref{fig:Diagrams}d, and is a continuous spectral function which is defined as~\cite{houtput2025}:
\begin{multline} \label{TomegaDef}
 \mathcal{T}_{ij}(\omega) \equiv \frac{e^2}{2\hbar \epsvac (2\pi)^3} \sum_{\nu_1 \nu_2} \int_{\text{1BZ}} \!\!\!\! \dee^3\mathbf{q} Y_{\nu_1\nu_2 i}(\mathbf{q}) Y^*_{\nu_1\nu_2 j}(\mathbf{q})  \\
 \times \Big\{\big[1+n(\hbar\omega_{\mathbf{q}\nu_1})+n(\hbar\omega_{\mathbf{q}\nu_2})\big]  \delta(\omega - |\omega_{\mathbf{q}\nu_1} + \omega_{\mathbf{q}\nu_2}|) \\
 +\big|n(\hbar\omega_{\mathbf{q}\nu_2})-n(\hbar\omega_{\mathbf{q}\nu_1})\big| \delta(\omega - |\omega_{\mathbf{q}\nu_1} - \omega_{\mathbf{q}\nu_2}|)\Big\}.
\end{multline}
In this expression, $Y_{\nu_1\nu_2 i}(\mathbf{q})$ 
%serves the same purpose as the mode polarity, in the sense that it 
represents the strength of the long-range interaction between an electron and two phonons, in branches $\nu_1, \nu_2$ and with approximately opposite momenta $-\mathbf{q}_1 \approx \mathbf{q}_2 = \mathbf{q}$. 
It is given in terms of the phonon frequencies $\omega_{\mathbf{q}\nu}$ and eigenvectors $e_{\kappa \alpha \nu}(\mathbf{q})$, and the derivative of the dynamical matrix with respect to the induced electric field $\frac{\partial \mathcal{D}_{\kappa \alpha,\kappa' \beta}(\mathbf{q})}{\partial \bm{\mathcal{E}}}$~\cite{houtput2025}:
\begin{multline}\label{Ydef}
    Y_{\nu_1\nu_2 i}(\mathbf{q}) \equiv \frac{1}{ie} \sqrt{\frac{\hbar}{2\omega_{\mathbf{q}\nu_1}} \frac{\hbar}{2\omega_{\mathbf{q}\nu_2}} }  \\
     \times \sum_{\kappa \alpha \kappa' \beta}  e^*_{\kappa \alpha \nu_1}(\mathbf{q}) \frac{\partial \mathcal{D}_{\kappa \alpha\kappa' \beta}(\mathbf{q})}{\partial \mathcal{E}_i}  e_{\kappa' \beta \nu_2}(\mathbf{q}). 
\end{multline}
In this manuscript, we focus on the cubic phase of CsPbI$_3$.
In this case, $\bm{\varepsilon}^{\infty}$, $\bm{\mathcal{P}}(\omega)$, and $\bm{\mathcal{T}}(\omega)$ all reduce to scalars \cite{houtput2025}:
\begin{align}
    \varepsilon^{\infty}_{ij} & = \varepsilon^{\infty} \delta_{ij}, \label{epsilonCubic} \\
    \mathcal{P}_{ij}(\omega) & = \mathcal{P}(\omega) \delta_{ij}, \label{PomegaCubic} \\
    \mathcal{T}_{ij}(\omega) & = \mathcal{T}(\omega) \delta_{ij}. \label{TomegaCubic}
\end{align}
Since these tensors only have a single independent component, it is sufficient to align the electric field $\bm{\mathcal{E}}$ along only one of the three cartesian axes. 
In this manuscript, we arbitrarily choose $\bm{\mathcal{E}}$ along the $z$-direction, meaning that we calculate $\mathcal{T}(\omega) \equiv \mathcal{T}_{zz}(\omega)$ by setting $i = j = z$ in \eref{TomegaDef}.

The second contribution is $\mathcal{R}(\omega)$, which is the Fr\"ohlich-type contribution from the linear electron-phonon interaction ~\cite{frohlich1954,miglio2020, guster2021}. It is a sum of delta functions at the longitudinal optical phonon frequencies $\omega_{\nu}$ at $\mathbf{q} = \mathbf{0}$~\cite{houtput2025}:
\begin{equation} \label{RomegaDef}
\mathcal{R}(\omega) \equiv \frac{e^2}{2 \epsvac \Omega_0} \sum_{\nu} \frac{|p_{\nu}|^2}{\omega_{\nu}} \delta(\omega-\omega_{\nu}),
\end{equation}
where $\Omega_0$ is the size of the unit cell and $p_{\nu}$ is the mode polarity, which represents the strength of the long-range interaction between an electron and one phonon branch $\nu$ with momentum $\mathbf{q}\approx\mathbf{0}$~\cite{houtput2025}.
Once both contributions are known, $\mathcal{P}(\omega)$ is simply evaluated as the sum of the two:
\begin{equation} \label{PomegaDef}
\mathcal{P}(\omega) = \mathcal{R}(\omega) + \mathcal{T}(\omega).
\end{equation}
We note that $\mathbf{q}$ is not restricted to $\mathbf{q}\approx \mathbf{0}$ in \erefs{TomegaDef}{Ydef}.
Therefore, to calculate the nonlinear spectral function $\mathcal{T}(\omega)$ using \eref{TomegaDef}, we need the phonon frequencies over the entire first BZ (see also \figref{fig:Inputs}b), despite making the long-range approximation~\cite{houtput2025}. 
This is counterintuitive because for the familiar linear electron-phonon interaction, only phonons with $\mathbf{q}\approx \mathbf{0}$ contribute to \eref{RomegaDef}. 

The $p_{\nu}$ in \eref{RomegaDef} can be written in terms of the Born effective charge tensor and phonon eigenvectors \cite{houtput2025, SIref}, and $Y_{\nu_1\nu_2z}(\mathbf{q})$ can be written in terms of phonon frequencies and eigenvectors as well as their derivative with respect to an external electric field~\cite{houtput2025}. 
All of these can be calculated from first principles (see \secref{app:DFTinputs} in the Supplemental Material ~\cite{SIref} and references \cite{kresse1996, kresse1996a, trots2008, togo2023, togo2023a, souza2002} therein), which allows to evaluate the self energy in \eref{SelfEnergy}.
Finally, in order to compute the electron mobility in \eref{muSERTA}, we need an expression for the electron bands $\varepsilon_{\mathbf{k}n}$. 
At sufficiently low carrier densities, only the electronic states close to the conduction band minimum will contribute.
Around the conduction band minimum, CsPbI$_3$ has a single parabolic conduction band:
\begin{equation} \label{ElectronBand}
\varepsilon_{\mathbf{k}} = \frac{\hbar^2 |\mathbf{k}|^2}{2m^*_e},
\end{equation}
where we take $m^*_e = 0.17 m_{\text{el}}$ for the effective band mass of the electron~\cite{ponce2019}.
We note that at low carrier densities, the hole mobility is calculated with a similar equation as \eref{muSERTA} for the electrons, except that we need to use the effective hole mass $m^*_h = 0.16 m_{\text{el}}$~\cite{ponce2019}. 
Ultimately, the effective band mass $m^*$ only enters the mobility as a prefactor according to $\mu \sim (m^*)^{-\frac{3}{2}}$ (see also \eref{muSERTApractical} in the Supplemental Material~\cite{SIref}). 
Therefore, we only need to calculate the electron mobility: the carrier mobility $\mu \equiv \mu_e + \mu_h$ can then be obtained as $\mu = \mu_e [1+(m^*_e/m^*_h)^{3/2}]$.

\figref{fig:Inputs} shows the material properties of cubic CsPbI$_3$ needed for the retarded self energy in \eref{SelfEnergy}, calculated  from first principles using the method described in \citeref{houtput2025}.
These material properties include the phonon frequencies $\omega_{\mathbf{q}\nu}$, the one-electron-two-phonon strengths $Y_{\nu_1\nu_2z}(\mathbf{q})$, and the electron-phonon spectral function $\mathcal{P}(\omega)  = \mathcal{R}(\omega) + \mathcal{T}(\omega)$. 
Additionally, we have calculated $\varepsilon^{\infty} = 5.96$ for the dielectric tensor of CsPbI$_3$, close to other first principles calculations in the literature \cite{ponce2019, su2021}.
The phonon calculations, with and without external electric field, were all performed in the harmonic approximation at zero temperature. 
This means that there are imaginary phonon modes in \figref{fig:Inputs}, since the cubic phase is only stabilized by anharmonic effects at finite temperatures~\cite{gu2021}. 
At lower temperatures, the tetragonal or orthorhombic phases are stable. 
However, due to the computational cost of a larger unit cell with an external electric field, we are limited to the cubic phase. 
To calculate the nonlinear electron-phonon spectral function $\mathcal{T}(\omega)$ in \figref{fig:Inputs}c, we set $Y_{\nu_1\nu_2z}(\mathbf{q}) = 0$ for all branches with frequencies $\omega_{\bm{q}\nu} < 0.1 ~\text{THz}$ in order to circumvent the imaginary branches. 
The value $0.1 ~\text{THz}$ is arbitrarily chosen; later, we show results with the cutoff varying from $0.05 ~\text{THz}$ to $0.2 ~\text{THz}$ to show that the qualitative conclusions are independent of this choice. 
It is not straightforward to determine how the imaginary modes would modify the nonlinear electron–phonon spectral function $\mathcal{T}(\omega)$ if they were treated within an anharmonic phonon theory.
Two competing effects are expected.
On the one hand, including additional phonon branches would increase $\mathcal{T}(\omega)$, since each branch contributes positively to \eref{TomegaDef}.
On the other hand, anharmonic renormalization generally shifts phonon frequencies upward, thereby reducing the Bose–Einstein occupation factors $n(\hbar\omega_{\mathbf{q}\nu})$ entering \eref{TomegaDef}.
The net effect on $\mathcal{T}(\omega)$ is therefore not known a priori.
Nevertheless, we expect the present calculation to capture the correct order of magnitude of the nonlinear electron–phonon contribution.

Importantly, the linear electron-phonon spectral function $\mathcal{R}(\omega)$ is independent of temperature, whereas the nonlinear electron-phonon spectral function $\mathcal{T}(\omega)$ depends on temperature through the Bose-Einstein distributions $n(\hbar\omega_{\mathbf{q}\nu})$. 
At sufficiently large temperatures $k_{\text{B}} T \gg \hbar \omega_{\mathbf{q}\nu}$, the Bose-Einstein distribution satisfies $n(\hbar\omega_{\mathbf{q}\nu}) \approx \frac{k_{\text{B}} T}{\hbar \omega_{\mathbf{q}\nu}} - \frac{1}{2}$ so that \eref{TomegaDef} scales linearly with temperature. 
CsPbI$_3$ has low phonon frequencies, so all phonons are thermally activated at room temperature. 
Additionally, \figref{fig:Inputs}b shows that most of the contributions to $\mathcal{T}(\omega)$ come from phonons with frequencies lower than $1 ~\text{THz}$. 
Therefore, we see a strong temperature dependence in \figref{fig:Inputs}c. 
Although $\mathcal{T}(\omega)$ is very small at zero temperature, it is of order $0.1$ at room temperature, which means one needs to look at finite temperature properties of CsPbI$_3$ in order to see the effect of nonlinear interaction. 
A similar result is seen for the temperature-dependent bandgap renormalization in methylammonium lead iodide (MAPbI$_3$)~\cite{saidi2016}.

\begin{figure}
\centering
\includegraphics[width=8.6cm]{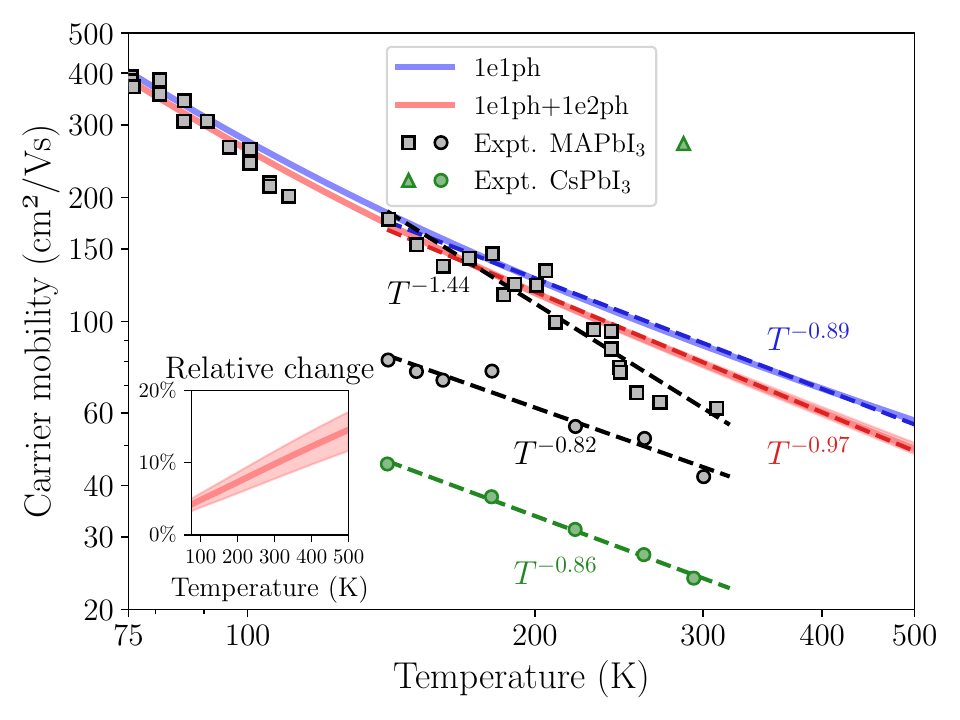}
\caption{\label{fig:Mobility} SERTA carrier mobility of CsPbI3 as a function of temperature, calculated using \eref{tauInvPractical} and \eref{muSERTApractical} of the Supplemental Material \cite{SIref} and the data from \figref{fig:Inputs}.
The solid blue line is calculated using only the linear electron-phonon interaction from \eref{RomegaDef}, and the solid red line is calculated including the one-electron-two-phonon interaction from \eref{TomegaDef}.
The inset shows the relative change of the mobility due to the one-electron-two-phonon interaction, relative to the mobility when nonlinear interaction is neglected.
Calculations are performed using a frequency cutoff of $0.1~\text{THz}$; the red shaded area shows the result when this cutoff is varied between $0.05~\text{THz}$ and $0.2~\text{THz}$, indicating that the nonlinear contribution remains significant regardless of this cutoff value.
Symbols indicate experimental mobility measurements in CsPbI3 (green) and MAPbI$_3$ (gray): squares denote single crystal THz conductivity measurements \cite{xia2021}, circles denote thin film time-resolved microwave conductivity measurements \cite{hutter2018}, and the triangle denotes a thin film optical pump measurement \cite{zhang2021}.
Dashed lines indicate power law fits in the region between $140-500~\text{K}$.
}
\end{figure}

With the electron-phonon spectral function $\mathcal{P}(\omega)$ from \figref{fig:Inputs}, we can calculate the SERTA carrier mobility from \erefs{muSERTA}{ElectronBand} (see also \secref{app:Mobility} in the Supplemental Material~\cite{SIref} and references \cite{cao2018, mahan2000} therein).
\figref{fig:Mobility} shows the calculated carrier mobility.
In line with the results of \figref{fig:Inputs}, the effects of nonlinear electron-phonon interaction are only revealed at higher temperatures.
Above room temperature, the electron mobility in CsPbI$_3$ is reduced by about $10\%$ by the long-range one-electron-two-phonon interaction.
Additionally, the temperature dependence of $\mathcal{T}(\omega)$ in \figref{fig:Inputs}c changes the temperature scaling of the mobility.
In the region between $140-500~\text{K}$, the mobility is described by a power law fit $\mu \sim T^{-\gamma}$, but the nonlinear electron-phonon interaction increases the exponent from $\gamma = 0.89$ to $\gamma = 0.97$. 

In a previous work \cite{houtput2025}, we have shown that the contribution of the nonlinear interaction to the polaron ground state energy  is only $0.013\%$ in the harmonic semiconductor LiF, and only $0.077\%$ in the anharmonic perovskite KTaO$_3$. 
The much larger, $\sim 10\%$ contribution found here for the mobility of CsPbI$_3$ arises from the combined action of three factors.
First, the strong anharmonicity of CsPbI$_3$ leads to large ionic displacements and therefore the nonlinear coupling matrix elements $Y_{\nu\nu'i}(\mathbf{q})$ in \eref{TomegaDef} are larger than in LiF or KTaO$_3$.
Second, mobility is a finite-temperature property, so thermal phonon populations amplify the nonlinear contribution through the Bose–Einstein factors $n(\hbar \omega_{\mathbf{q}\nu})$ in \eref{TomegaDef}.
Third, the low phonon frequencies associated with the heavy constituent atoms of CsPbI$_3$ result in large phonon occupations at room temperature over a broad range of modes.
The nonlinear contribution is therefore expected to become appreciable when strong anharmonicity, finite-temperature phonon populations, and low-energy phonons occur simultaneously.
Conversely, if one or more of these conditions is absent, as in LiF or KTaO$_3$, nonlinear electron–phonon effects are expected to remain small.
This provides a simple physical criterion for identifying materials and finite-temperature properties for which nonlinear electron–phonon interactions may need to be explicitly included, in particular in anharmonic halide perovskites.

To assess both the strengths and limitations of our long-range model, we compare our results with a first-principles calculation including both long- and short-range linear electron–phonon interactions (\figref{fig:CsPbI3_LRbenchmark} of the Supplemental Material \cite{SIref}) and with experimental mobility measurements for CsPbI$_3$ and methylammonium lead iodide (MAPbI$_3$) \cite{xia2021, hutter2018, zhang2021}.
The mobility of MAPbI$_3$ provides a useful reference because it is likewise largely governed by Pb and I displacements \cite{fu2018, hutter2018, ponce2019}.
As shown in \figref{fig:Mobility}, the measured mobilities exhibit considerable scatter; we therefore first compare their temperature dependence through the power-law relation $\mu \sim T^{-\gamma}$.
The linear long-range model yields $\gamma = 0.89$, in good agreement with thin-film measurements ($\gamma = 0.82$; $\gamma = 0.86$).
However, transport in polycrystalline thin films can be strongly affected by grain boundaries, which are not included in our model.
A more direct comparison is therefore provided by single-crystal measurements.
In this case, the absolute mobility values are comparable to our calculations, whereas the experimental temperature exponent, $\gamma = 1.44$, is larger than predicted by the linear long-range model.
Notably, including the nonlinear electron–phonon interaction shifts the calculated temperature dependence toward the experimental value.
In \secref{app:LRbenchmarking} of the Supplemental Material \cite{SIref}, we further compare the linear long-range model with a full first-principles calculation using the complete linear matrix elements $g_{mn\nu}(\mathbf{k},\mathbf{q})$, evaluated on $144 \times 144 \times 144$ $\mathbf{k}$-grid and $\mathbf{q}$-grid.
The full calculation gives $\gamma = 1.16$, confirming that the long-range approximation does not fully reproduce the temperature scaling, while yielding similar absolute mobility values.
Thus, the approximations adopted here provide a reasonable estimate of the mobility magnitude and, more importantly, do not alter our central conclusion that nonlinear electron–phonon interactions reduce the mobility by approximately $10\%$.
The remaining discrepancy with the experiment is likely due to the difference between the cubic and orthorhombic phases, the neglect of anharmonicity and one-electron-$n$-phonon nonlinear interaction terms with $n\geq 3$, the long-range approximation, or the use of the SERTA.

The present work suggests many perspectives for further research, several of which consist of improving the approximations made in this Letter.
These perspectives can be separated into two distinct categories: improving the calculation of $\mathcal{P}(\omega)$, and using a more refined theory for electronic transport. 
Firstly, the calculation of $\mathcal{P}(\omega)$ can be improved by including phonon anharmonicity and all orders of one-electron-$n$-phonon interactions. 
Most notably, the results in \figrefs{fig:Inputs}{fig:Mobility} are based on the harmonic approximation for the phonons, but CsPbI$_3$ is known to exhibit strong phonon anharmonicity \cite{gu2021}.
Since higher-order nonlinear interactions will involve higher powers of the Bose-Einstein occupation factors, they are expected to scale even faster with temperature than the linear and one-electron-two-phonon interaction.
Depending on the prefactors associated with each of these terms, this scaling may significantly affect the power law scaling of electron mobility at room temperatures, similarly to the result that \figref{fig:Mobility} shows for the one-electron-two-phonon interaction.

The second category of perspectives is to take any result for the long-range electron-phonon spectral function $\mathcal{P}(\omega)$ as given, and use it to calculate the electron mobility with more advanced transport theories.
One may start from the self energy in \eref{SelfEnergy}, written in terms of $\mathcal{P}(\omega)$ and the harmonic quantities $\varepsilon_{\mathbf{k}n}$ and $\varepsilon^{\infty}$.
This expression can be used instead of the Fan-Migdal self energy to derive the equivalent Boltzmann transport equation \cite{ponce2020}, or even more general transport theories that do not make the quasiparticle approximation \cite{lihm2025,lihm2026}.
This may significantly affect the results for the mobility, even given the same spectral function $\mathcal{P}(\omega)$.

We have shown that carrier mobility in CsPbI$_3$ receives a sizeable contribution from nonlinear one-electron–two-phonon interactions, which are commonly neglected in transport calculations.
This contribution arises from the combined effects of strong lattice anharmonicity, low phonon frequencies, and the resulting large thermal phonon populations at finite temperature.
These ingredients are characteristic of halide perovskites and, more generally, of soft polar anharmonic materials, suggesting that nonlinear electron–phonon effects may be relevant well beyond CsPbI$_3$.
Our results therefore motivate further theoretical developments capable of treating nonlinear electron–phonon interactions beyond the present approximations, including anharmonic phonon renormalization, higher-order one-electron–$n$-phonon processes, and short-range coupling.
The framework introduced here and in \citeref{houtput2025} provides a first step toward such a more complete description.

\section*{Data availability statement}
The first-principles data and postprocessing code used to generate \figrefs{fig:Inputs}{fig:Mobility} are publicly available online \cite{1e2ph-spectral, MCA-houtput2026}.

\acknowledgments

M.H., I.Z., W.W., S.K. and J.T. acknowledge the Research Foundation Flanders (FWO), file numbers 1224724N, V472923N, 1120825N, G060820N, G0A9F25N, and G0AIY25N, for their funding of this research.
C.F. and J.T. acknowledge support from the joint Austrian Science Fund (FWF) - FWO project I 4506.
S. P. is a Research Associate of the Fonds de la Recherche Scientifique - FNRS.
This publication was supported by the Walloon Region in the strategic axe FRFS-WEL-T.

\bibliography{References}

% We manually include Supplemental_Material.aux so the cross-referencing works

\end{document}

\title{
Supplemental Material for ``Importance of nonlinear long-range electron-phonon interaction for the carrier mobility of anharmonic halide perovskites''
}

\author{Matthew Houtput}
\affiliation{Theory of Quantum Systems and Complex Systems, Universiteit Antwerpen, B-2000 Antwerpen, Belgium}

\author{Ingvar Zappacosta}
\affiliation{Theory of Quantum Systems and Complex Systems, Universiteit Antwerpen, B-2000 Antwerpen, Belgium}

\author{William Wenborn}
\affiliation{Theory of Quantum Systems and Complex Systems, Universiteit Antwerpen, B-2000 Antwerpen, Belgium}

\author{Serghei Klimin}
\affiliation{Theory of Quantum Systems and Complex Systems, Universiteit Antwerpen, B-2000 Antwerpen, Belgium}

\author{Samuel Ponc\'e}
\affiliation{European Theoretical Spectroscopy Facility and Institute of Condensed Matter and Nanosciences, Universit\'e catholique de Louvain, Chemin des \'Etoiles 8, B-1348 Louvain-la-Neuve, Belgium}
\affiliation{WEL Research Institute, avenue Pasteur 6, 1300 Wavre, Belgium}

\author{Jacques Tempere}
\affiliation{Theory of Quantum Systems and Complex Systems, Universiteit Antwerpen, B-2000 Antwerpen, Belgium}

\author{Cesare Franchini}
\affiliation{Faculty of Physics, Computational Materials Physics, University of Vienna, Kolingasse 14-16, Vienna A-1090, Austria}
\affiliation{Department of Physics and Astronomy, Alma Mater Studiorum - Universit\`a di Bologna, Bologna, 40127 Italy}

\maketitle

\section{Mobility in the SERTA} \label{app:Mobility}
In the main manuscript, we have described how to calculate the electron mobility in the SERTA using \erefs{muSERTA}{SelfEnergy}.
While these equations describe the physical intuition behind our method, they are not yet ready for a practical numerical evaluation.
In this section, we provide the further analytical steps and approximations that lead to the expression for the mobility that we evaluate in practice.
The chain of reasoning is very similar to the one presented in the supplementary information of~\cite{ponce2019}.

We start from the following, general expression for the on-shell Fan-Migdal self energy in the diagonal band approximation, in terms of the two-argument Eliashberg function $\alpha^2 F_{\mathbf{k},n}(\varepsilon, \omega)$~\cite{giustino2017}:
\begin{equation} \label{FanMigdalGeneral}
    \Sigma^{\rm R}_{\mathbf{k}nn}\left(\varepsilon \right) = \hbar \int_{-\infty}^{+\infty} \!\!\!\! \dee \varepsilon' \int_{0}^{+\infty} \!\!\!\! \dee \omega \ \alpha^2 F_{\mathbf{k}nn}(\varepsilon', \omega) \left[ \frac{1 - f^0(\varepsilon') + n(\hbar\omega)}{\varepsilon - \varepsilon' - \hbar \omega + i\delta} + \frac{f^0(\varepsilon') + n(\hbar\omega)}{\varepsilon - \varepsilon' + \hbar \omega + i\delta} \right].
\end{equation}
Within the scope of this article, we are interested in the Eliashberg function calculated using only dipole long-range electron-phonon interaction.
In~\cite{houtput2025}, it is shown that the Eliashberg function in this case is given by the following expression:

\begin{equation} \label{EliashbergLongRange}
    \alpha^2 F_{\mathbf{k}nn}(\varepsilon', \omega) \approx \frac{e^2}{\varepsilon_0 (2\pi)^3} \int_{\mathbb{R}^3} \!\! \dee^3 \mathbf{Q} 
        \frac{\mathbf{Q} \cdot \bm{\mathcal{P}}(\omega) \cdot \mathbf{Q} }{(\mathbf{Q}\cdot \bm{\varepsilon}^{\infty} \cdot \mathbf{Q})^2}
    \delta(\varepsilon' - \varepsilon_{\mathbf{k}+\mathbf{Q}n}),
\end{equation}
Combining \erefs{FanMigdalGeneral}{EliashbergLongRange}, and evaluating the self energy on-shell by setting $\varepsilon = \varepsilon_{\mathbf{k}n}$, yields Eq.~(4) from the main manuscript.
We reproduce it here for clarity:
\begin{multline}\label{SelfEnergy2}
\Sigma^{\rm R}_{\mathbf{k}nn}\left(\varepsilon_{\mathbf{k}n}\right) = \frac{\hbar e^2}{\epsvac (2\pi)^3} \int_{0}^{+\infty} \!\!\!\! \dee \omega \int_{\mathbb{R}^3} \!\! \dee^3 \mathbf{Q} \frac{\mathbf{Q} \cdot \bm{\mathcal{P}}(\omega) \cdot \mathbf{Q} }{(\mathbf{Q}\cdot \bm{\varepsilon}^{\infty} \cdot \mathbf{Q})^2} \\
\times \bigg[ \frac{1 - f^0(\varepsilon_{\mathbf{k}+\mathbf{Q}n}) + n(\hbar\omega)}{\varepsilon_{\mathbf{k}n} - \varepsilon_{\mathbf{k}+\mathbf{Q}n} - \hbar \omega + i\delta} + \frac{f^0(\varepsilon_{\mathbf{k}+\mathbf{Q}n}) + n(\hbar\omega)}{\varepsilon_{\mathbf{k}n} - \varepsilon_{\mathbf{k}+\mathbf{Q}n} + \hbar \omega + i\delta} \bigg]. 
\end{multline}
For the rest of the derivation, we will limit ourselves to the case of cubic symmetry.
In this case, both of the tensors $\bm{\varepsilon}^{\infty}$ and $\bm{\mathcal{P}}(\omega)$ are reduced to scalars \cite{houtput2025}:
\begin{align}
    \varepsilon^{\infty}_{ij} & = \varepsilon^{\infty} \delta_{ij}, \\
    \mathcal{P}_{ij}(\omega) & = \mathcal{P}(\omega) \delta_{ij}.
\end{align}
With this assumption of cubic symmetry, \eref{SelfEnergy2} for the on-shell retarded self energy reduces to:
\begin{multline} \label{SelfEnergyCubic}
\Sigma^{\rm R}_{\mathbf{k}nn}\left(\varepsilon_{\mathbf{k}n}\right) = \frac{\hbar e^2}{\epsvac (\varepsilon^{\infty})^2 (2\pi)^3} \int_{0}^{+\infty} \!\!\!\! \dee \omega \ \mathcal{P}(\omega) \int_{\mathbb{R}^3} \frac{\dee^3 \mathbf{Q}}{|\mathbf{Q}|^2} \\
\times \bigg[ \frac{1 - f^0(\varepsilon_{\mathbf{k}+\mathbf{Q}n}) + n(\hbar\omega)}{\varepsilon_{\mathbf{k}n} - \varepsilon_{\mathbf{k}+\mathbf{Q}n} - \hbar \omega + i\delta} + \frac{f^0(\varepsilon_{\mathbf{k}+\mathbf{Q}n}) + n(\hbar\omega)}{\varepsilon_{\mathbf{k}n} - \varepsilon_{\mathbf{k}+\mathbf{Q}n} + \hbar \omega + i\delta} \bigg]. 
\end{multline}
The theory of \cite{houtput2025} then states that $\mathcal{P}(\omega) = \mathcal{R}(\omega) + \mathcal{T}(\omega)$ is a dimensionless electron-phonon spectral function, which gets contributions from the one-electron-one-phonon and one-electron-two-phonon interactions.

We will now make two further assumptions.
Firstly, we will assume that we are working with a single parabolic band, given by:
\begin{equation} \label{Econd}
\varepsilon_{\mathbf{k}} = \pm \frac{\hbar^2 |\mathbf{k}|^2}{2m^*}.
\end{equation}
where the top sign is for electrons in a conduction band, and the bottom sign is for holes in a valence band.
Without loss of generality, we adopt the convention that $\mathbf{k}=\mathbf{0}$ lies at the conduction band minimum (or the valence band maximum in the case of holes), even if that extremum does not occur at the $\bm{\Gamma}$-point. We will drop the index $n$ everywhere because we only have a single band.
Secondly, we will assume that the band gap $\Delta E$ of the material satisfies $\exp\left(\Delta E / 2 k_{\text{B}} T\right) \gg 1$. 
In this article, we focus on cubic CsPbI$_3$ which has a band gap of $\Delta E = 1.7~\text{eV}$~\cite{cao2018}.
We also limit ourselves to temperatures below $500~\text{K}$, which corresponds to $k_{\text{B}} T = 0.043 ~\text{eV}$, so this approximation is excellent.
In that case, we have $f^0(\varepsilon_{\mathbf{k}}) \approx 0$ for electrons, and $f^0(\varepsilon_{\mathbf{k}}) \approx 1$ for holes.
Then, we get the following expression for the Fan-Migdal self energy:
\begin{multline}
 \Sigma^{\rm R}_{\mathbf{k}}\left(\varepsilon_{\mathbf{k}}\right) = \lim_{\delta \rightarrow 0^+} \mp\frac{\hbar e^2}{\epsvac (\varepsilon^{\infty})^2 (2\pi)^3} \int_{0}^{+\infty} \!\!\!\! \dee \omega \ \mathcal{P}(\omega)  \\
\times \bigg\{ [1 + n_B(\omega)] \int_{\mathbb{R}^3} \frac{\dee^3 \mathbf{Q}}{|\mathbf{Q}|^2} \frac{1}{\frac{\hbar^2 |\mathbf{k}+\mathbf{Q}|^2}{2m^*} - \frac{\hbar^2 |\mathbf{k}|^2}{2m^*} + \hbar \omega \mp i\delta} \\
+ n_B(\omega) \int_{\mathbb{R}^3} \frac{\dee^3 \mathbf{Q}}{|\mathbf{Q}|^2} \frac{1}{\frac{\hbar^2 |\mathbf{k}+\mathbf{Q}|^2}{2m^*} - \frac{\hbar^2 |\mathbf{k}|^2}{2m^*} - \hbar \omega \mp i\delta} \bigg\}.
\end{multline}
The remaining integrals over $\mathbf{Q}$ can be calculated analytically.
The general form of this integral also appears in the calculation of the quasiparticle energy of the Fr\"ohlich polaron, so its result is known~\cite{mahan2000}:
\begin{equation} \label{arcsinIntegral}
\int_{\mathbb{R}^3} \frac{\dee^3 \mathbf{Q}}{|\mathbf{Q}|^2} \frac{1}{\frac{\hbar^2 |\mathbf{k}+\mathbf{Q}|^2}{2m^*} + C} = \frac{2\pi^2}{|\mathbf{k}|} \frac{2m^*}{\hbar^2} \arcsin\left(\sqrt{\frac{\frac{\hbar^2 |\mathbf{k}|^2}{2m^*}}{\frac{\hbar^2 |\mathbf{k}|^2}{2m^*}+C}}\right), \hspace{20pt} \text{ where } C \in \mathbb{C} \backslash \mathbb{R}^-_0.
\end{equation}
Therefore, the Fan-Migdal self energy is given by:
\begin{multline}
\Sigma^{\rm R}_{\mathbf{k}}\left(\varepsilon_{\mathbf{k}}\right) = \lim_{\delta \rightarrow 0^+} \mp\frac{e^2}{4\pi\epsvac (\varepsilon^{\infty})^2} \frac{2m^*}{\hbar |\mathbf{k}|} \int_{0}^{+\infty} \!\!\!\! \dee \omega \ \mathcal{P}(\omega) \\
\times \left( [1 + n_B(\omega)] \arcsin\left(\sqrt{\frac{\frac{\hbar |\mathbf{k}|^2}{2m^* \omega}}{1 \mp i \delta}}\right) + n_B(\omega) \arcsin\left(\sqrt{\frac{\frac{\hbar |\mathbf{k}|^2}{2m^*\omega}}{-1 \mp i \delta}}\right) \right) .
\end{multline}
To calculate the inverse lifetimes, we need to take the imaginary part of this expression.
Using the following identities, where $a>0$ and $\Theta(a)$ is the Heaviside step function:
\begin{align}
\text{Im}\left[\lim_{\delta \rightarrow 0^+} \arcsin\left(\sqrt{\frac{a}{1 \mp i \delta}}\right) \right] & = \pm\arccosh(\sqrt{a}) \Theta(a-1), \\
\text{Im}\left[\lim_{\delta \rightarrow 0^+} \arcsin\left(\sqrt{\frac{a}{-1 \mp i \delta}}\right) \right] & = \pm\arcsinh(\sqrt{a}),
\end{align}
we find the following expression for the inverse lifetimes:
\begin{align}
\frac{1}{\tau_{\mathbf{k}}} & = - \frac{2}{\hbar} \text{Im}\left[\Sigma^{\rm R}_{\mathbf{k}}\left(\varepsilon_{\mathbf{k}}\right) \right], \\
& =
\frac{2}{\hbar} \frac{e^2}{4\pi\epsvac (\varepsilon^{\infty})^2} \sqrt{\frac{2m^*}{\hbar}} \int_{0}^{+\infty} \!\!\!\! \dee \omega \frac{\mathcal{P}(\omega)}{\sqrt{\omega}} \varphi\left(\frac{\hbar^2 |\mathbf{k}|^2}{2 m^* k_{\text{B}} T}, \frac{\hbar \omega}{k_{\text{B}} T} \right). \label{tauK_SERTA}
\end{align}
Here, the auxiliary function $\varphi(x,x_0)$ is defined as follows~\cite{ponce2019}:
\begin{equation} \label{phiDef}
\varphi(x,x_0) := \sqrt{\frac{x_0}{x}} \left( \frac{\arccosh\left(\sqrt{\frac{x}{x_0}}\right) \Theta(x-x_0)}{1-e^{-x_0}} + \frac{\arcsinh\left(\sqrt{\frac{x}{x_0}}\right) }{e^{x_0}-1} \right).
\end{equation}
Note that \erefs{tauK_SERTA}{phiDef} no longer contain a top and bottom sign: this means these expressions are the same for electrons and for holes.
The only difference is that we should use the conduction band mass $m^*_e$ for electrons, and the valence band mass $m^*_h$ for holes.

The scattering time $\tau_{\mathbf{k}}$ is an isotropic function: it does not depend on the direction of $\mathbf{k}$.
To highlight this, in~\cite{ponce2019} the inverse scattering times are written as $\tau_{\mathbf{k}} = \tau\left(\frac{\hbar^2 |\mathbf{k}|^2}{2m^* k_{\text{B}} T}\right)$, where $\tau(x)$ is given by:
\begin{equation} \label{tauInvPractical}
\frac{1}{\tau(x)} = \frac{2}{\hbar} \frac{e^2}{4\pi\epsvac (\varepsilon^{\infty})^2} \sqrt{\frac{2m^*}{\hbar}} \int_{0}^{+\infty} \!\!\!\! \dee \omega \frac{\mathcal{P}(\omega)}{\sqrt{\omega}} \varphi\left(x, \frac{\hbar \omega}{k_{\text{B}} T} \right).
\end{equation}

With the expression for the electron lifetimes known, the SERTA electron mobility can be evaluated using \erefs{muSERTA}{ooeoF} from the main part of the manuscript \cite{ponce2019}, given here for a single conduction band:
\begin{equation} \label{muSERTA2}
\mu_{e,\alpha \beta} \approx -\frac{e}{n_e (2\pi)^3} \int_{\text{1BZ}} \!\!\!\! \dee^3\mathbf{k} \frac{\partial f^0(\varepsilon_{\mathbf{k}})}{\partial \varepsilon_{\mathbf{k}}} v_{\mathbf{k}\alpha} v_{\mathbf{k}\beta} \tau_{\mathbf{k}} . 
\end{equation}
We require two more elements in this expression: the band velocities $v_{\mathbf{k}\alpha}$, and the quantity $\frac{1}{n_e} \frac{\partial f^0(\varepsilon_{\mathbf{k}})}{\partial \varepsilon_{\mathbf{k}}}$.
For a parabolic band given by \eref{Econd}, the band velocities are easily calculated from the definition:
\begin{align}
v_{\mathbf{k}\alpha} & = \frac{1}{\hbar} \frac{\partial \varepsilon_{\mathbf{k}}}{\partial k_{\alpha}}, \nonumber \\
& = \frac{\hbar k_{\alpha}}{m^*}.
\end{align}
For the quantity $\frac{1}{n_e} \frac{\partial f^0(\varepsilon_{\mathbf{k}})}{\partial \varepsilon_{\mathbf{k}}}$ we will once again assume that $\Delta E \gg 2 k_{\text{B}} T$, so the Fermi-Dirac distribution is well approximated by a Boltzmann distribution:
\begin{align}
f^0(\varepsilon) & \approx \exp\left(\frac{\mu}{k_{\text{B}} T}\right) \exp\left(-\frac{\varepsilon}{k_{\text{B}} T}\right), \\
\Rightarrow \frac{\partial f^0(\varepsilon)}{\partial \varepsilon} & \approx -\frac{1}{k_{\text{B}} T}\exp\left(\frac{\mu}{k_{\text{B}} T}\right) \exp\left(-\frac{\hbar^2 |\mathbf{k}|^2}{2m^* k_{\text{B}} T}\right).
\end{align}
We may also calculate the electron density under this approximation:
\begin{align}
n_e & = \frac{1}{(2\pi)^3} \int_{\text{1BZ}} \!\!\!\! \dee^3\mathbf{k} f^0(\varepsilon_{\mathbf{k}}) , \\
& \approx \exp\left(\frac{\mu}{k_{\text{B}} T}\right) \frac{1}{(2\pi)^3} \int_{\mathbb{R}^3} \!\! \dee^3\mathbf{k} \exp\left(-\frac{\hbar^2 |\mathbf{k}|^2}{2m^* k_{\text{B}} T}\right), \\
& = \exp\left(\frac{\mu}{k_{\text{B}} T}\right) \frac{1}{8\pi^{\frac{3}{2}}} \left(\frac{2m^* k_{\text{B}} T}{\hbar^2} \right)^{\frac{3}{2}}.
\end{align}
To go from the first to the second line, we assumed that the integral is converged before $|\mathbf{k}|$ reaches the edge of the first Brillouin zone, so the integration domain can be approximately replaced by $\mathbb{R}^3$.
We find that $\frac{1}{n_e} \frac{\partial f^0(\varepsilon_{\mathbf{k}})}{\partial \varepsilon_{\mathbf{k}}}$ is independent of the chemical potential:
\begin{equation} \label{f0ne}
\frac{1}{n_e} \frac{\partial f^0(\varepsilon_{\mathbf{k}})}{\partial \varepsilon_{\mathbf{k}}} = -\frac{8\pi^{\frac{3}{2}}}{k_{\text{B}} T}  \left(\frac{\hbar^2}{2m^* k_{\text{B}} T} \right)^{\frac{3}{2}} \exp\left(-\frac{\hbar^2 |\mathbf{k}|^2}{2m^* k_{\text{B}} T}\right)
\end{equation}
The presented reasoning was for electrons in the conduction band.
For holes in the valence band, we would have to approximate the hole distribution function $1-f^0(\varepsilon)$ by a Boltzmann distribution.
Then, a similar reasoning shows that \eref{f0ne} also holds for holes, with a hole density $n_h$.
Plugging \eref{f0ne} back into \eref{muSERTA2} for the electron mobility yields:
\begin{equation}
\mu_{e,\alpha \beta} = \frac{e}{m^*} \frac{2}{\pi^{\frac{3}{2}}} \left(\frac{\hbar^2}{2m^* k_{\text{B}} T} \right)^{\frac{5}{2}} \int_{\mathbb{R}^3} \dee^3\mathbf{k} \ k_{\alpha}  k_{\beta} \exp\left(-\frac{\hbar^2 |\mathbf{k}|^2}{2m^* k_{\text{B}} T}\right) \tau\left(\frac{\hbar^2 |\mathbf{k}|^2}{2m^* k_{\text{B}} T}\right) .
\end{equation}
This integral is most easily evaluated in spherical coordinates, $\mathbf{k} = k \mathbf{n}$ where $k \in [0,+\infty[$ and $\mathbf{n} \in S^2$ is a direction vector on the unit sphere.
Then, the integral factorizes in a radial and angular part:
\begin{equation}
\mu_{e,\alpha \beta} = \frac{e}{m^*} \frac{2}{\pi^{\frac{3}{2}}} \left(\frac{\hbar^2}{2m^* k_{\text{B}} T} \right)^{\frac{5}{2}} \int_{S^2} d^2 \mathbf{n} \ n_{\alpha} n_{\beta} \int_0^{+\infty} \!\! \dee k \ k^4  \exp\left(-\frac{\hbar^2 k^2}{2m^* k_{\text{B}} T}\right) \tau\left(\frac{\hbar^2 k^2}{2m^* k_{\text{B}} T}\right) .
\end{equation}
The angular integral can be calculated explicitly.
From symmetry it can be seen that it is zero when $\alpha \neq \beta$, and that result does not depend on the value of $\alpha$ when $\alpha = \beta$. Therefore, we only need to calculate the case $\alpha = \beta = z$:
\begin{align}
    \int_{S^2} d^2 \mathbf{n} \ n_{\alpha} n_{\beta} & = \int_{S^2} d^2 \mathbf{n} \ n^2_{z} \ \delta_{\alpha \beta}, \\
    & = \int_{0}^{\pi} \dee \theta \int_{0}^{2\pi} \dee \phi \cos^2(\theta) \sin(\theta) \ \delta_{\alpha \beta}, \\
    & = \frac{4\pi}{3} \delta_{\alpha \beta}
\end{align}
This means the mobility tensor can be written as $\mu_{\alpha \beta} = \mu \delta_{\alpha \beta}$: it is isotropic, as expected for a cubic material.
The isotropic mobility is given by:
\begin{equation}
\mu_e = \frac{e}{m^*} \frac{8}{3\sqrt{\pi}} \left(\frac{\hbar^2}{2m^* k_{\text{B}} T} \right)^{\frac{5}{2}} \int_0^{+\infty} \!\! \dee k \ k^4 \exp\left(-\frac{\hbar^2 k^2}{2m^* k_{\text{B}} T}\right) \tau\left(\frac{\hbar^2 k^2}{2m^* k_{\text{B}} T}\right).
\end{equation}
Finally, we make the substitution $x := \frac{\hbar^2 k^2}{2m^* k_{\text{B}} T}$ to arrive at our practical expression for the SERTA:
\begin{equation}\label{muSERTApractical}
    \mu_e = \frac{e}{m^*} \left[\frac{4}{3\sqrt{\pi}} \int_0^{+\infty} \!\! \dee x \ \tau(x) x^{\frac{3}{2}} e^{-x} \right].
\end{equation}
To evaluate $\tau(x)$ in practice, we use expression \eref{tauInvPractical}, where we use $\mathcal{P}(\omega) = \mathcal{R}(\omega) + \mathcal{T}(\omega)$ as defined by \eref{TomegaDef} and \erefs{RomegaDef}{PomegaDef}:
\begin{multline}
    \frac{1}{\tau(x)} = \frac{1}{\hbar \Omega_0} \frac{e^4}{4\pi \epsvac^2 (\varepsilon^{\infty})^2} \sqrt{\frac{2m^*}{\hbar}} \sum_{\nu \in \text{LO}} \frac{|p_{\nu}|^2}{\omega_{\nu}^{\frac{3}{2}}} \varphi\left(x, \frac{\hbar \omega_{\nu}}{k_{\text{B}} T} \right) \\
     +  \frac{2}{\hbar} \frac{e^2}{4\pi\epsvac (\varepsilon^{\infty})^2} \sqrt{\frac{2m^*}{\hbar}} \int_{0}^{+\infty} \!\!\!\! \dee \omega \frac{\mathcal{T}(\omega)}{\sqrt{\omega}} \varphi\left(x, \frac{\hbar \omega}{k_{\text{B}} T} \right), \label{tauSERTApractical}
\end{multline}
where we recall that $\varphi\left(x, x_0 \right)$ is simply the analytical function given in \eref{phiDef}. 
Additionally, $p_{\nu}$ is the mode polarity, which can be calculated from first principles since it is given in terms of the Born effective charge tensor $\mathbf{Z}_{\kappa}$, the phonon eigenvectors $\mathbf{e}_{\kappa, \nu}(\mathbf{q})$ at $\Gamma$, and the ionic masses $m_{\kappa}$ \cite{houtput2025}:
\begin{equation}
    p_{\nu} \equiv \lim_{\mathbf{q} \rightarrow \mathbf{0}} \sum_{\kappa} \frac{\mathbf{q}\cdot \mathbf{Z}_{\kappa} \cdot \mathbf{e}_{\kappa \nu}(\mathbf{q})}{|\mathbf{q}| \sqrt{m_{\kappa}}}
\end{equation}
For a given temperature and electron-phonon spectral function $\mathcal{T}(\omega)$, expression \eref{tauSERTApractical} is evaluated through numerical integration.
The result is used in \eref{muSERTApractical} to calculate the mobility through one more numerical integration. 

\section{DFT input parameters} \label{app:DFTinputs}
\subsection{Electronic parameters} \label{appsec:DFTelectronic}
Our calculations are performed in VASP~\cite{kresse1996, kresse1996a} with the Perdew-Burke-Ernzerhof (PBE) functional.
We choose $a = 6.276~\text{\AA}$ for the lattice parameter as in~\cite{ponce2019}, which is close to the experimental lattice parameter of $a = 6.289~\text{\AA}$~\cite{trots2008}.
For our pseudopotentials, we choose GW pseudopotentials from the VASP potpaw.64 dataset, in order to have access to pseudopotentials that contain the 5s$^2$ and 5p$^6$ states for lead.
Specifically, we choose the following pseudopotentials: Cs\_sv\_GW [5s$^2$ 5p$^6$ 6s$^1$] for cesium, Pb\_sv\_GW [5s$^2$ 5p$^6$ 5d$^{10}$ 6s$^2$ 6p$^2$] for lead, and I\_GW [5s$^2$ 5p$^5$] for iodine.
The plane wave cutoff energy was set to $600~\text{eV}$.
Unit cell properties are calculated on a $\mathbf{k}$-grid of size $8 \times 8 \times 8$.
For phonon properties calculated on a $2 \times 2 \times 2$ supercell, an appropriately scaled $\mathbf{k}$-grid of size $4 \times 4 \times 4$ is used.
The Born effective charge tensor $\mathbf{Z}_{\kappa}$ and dielectric tensor $\bm{\varepsilon}^{\infty}$ are calculated from density functional perturbation theory (DFPT) with a $16 \times 16 \times 16$ $\mathbf{k}$-grid.
All results in the manuscript are calculated without spin-orbit coupling.

\subsection{Phonon parameters} \label{appsec:DFTphonon}
As in~\cite{houtput2025}, the phonon properties are calculated using the finite displacement method implemented in the PhonoPy package~\cite{togo2023, togo2023a}.
The phonon frequencies must be converged with respect to the supercell size, and the magnitude $\Delta x$ of the ion displacements.
We found that the phonon frequencies are more or less converged after choosing $\Delta x = 0.005~\text{\AA}$: more motivation for this value is provided later in this section.
We will show in a moment that the calculations for the nonlinear electron-phonon interaction require strict convergence tolerances, making the calculations computationally expensive.
Therefore, we are limited to small supercells, and as such, all phonon results in this manuscript are computed on a $2 \times 2 \times 2$ supercell.

\subsection{Parameters for $\mathcal{T}(\omega)$}
The evaluation of the one-electron-two-phonon spectral function $\mathcal{T}(\omega)$ through \eref{TomegaDef} requires an integral over Dirac delta functions over the first Brillouin zone. 
As in~\cite{houtput2025}, we calculate these integrals using the smearing method.
We replace the Dirac delta with a Gaussian with a finite width $\sigma = 0.025~\text{THz}$:
\begin{equation}
    \delta(\omega) \rightarrow \frac{1}{\sqrt{2\pi} \sigma} e^{-\frac{\omega^2}{2 \sigma^2}}
\end{equation}
Then, the integral over the first Brillouin zone is evaluated using the trapezoid rule on a fine $\mathbf{q}$-grid.
All quantities are interpolated from the coarse $2 \times 2 \times 2$ commensurate grid to the fine grid by Fourier interpolation~\cite{houtput2025}.
We use a $128 \times 128 \times 128$ fine $\mathbf{q}$-grid for all results of the electron-phonon spectral function, and a $32 \times 32 \times 32$ fine $\mathbf{q}$-grid for all results of the electron lifetime and mobility.

For the calculation of the electric field derivative of the dynamical matrix $\frac{\partial \mathcal{D}_{\kappa \alpha, \kappa' \beta}(\mathbf{q})}{\partial \mathcal{E}_z}$ that appears in Eqs.~(8)-(9) of the main manuscript, there are several numerical parameters which we must pay special attention to.
We again follow~\cite{houtput2025}, where the derivative of the dynamical matrix is approximated by a finite difference:
\begin{equation} \label{FiniteDifference}
\frac{\partial \mathcal{D}_{\kappa \alpha, \kappa' \beta}(\mathbf{q})}{\partial \mathcal{E}_z} \approx \frac{\mathcal{D}_{\kappa \alpha, \kappa' \beta}(\mathbf{q}, \mathcal{E}_z) - \mathcal{D}_{\kappa \alpha, \kappa' \beta}(\mathbf{q}, - \mathcal{E}_z)}{2\mathcal{E}_z} + O(\mathcal{E}_z^3)
\end{equation}
We need to calculate the dynamical matrix from first principles, using a calculation with an appropriate value of the external electric field $\mathcal{E}_z$.
The main issue in practice is if we choose $\mathcal{E}_z$ too small, the change in the dynamical matrix may be smaller than the numerical noise that is introduced during the calculation.
Therefore, we should choose $\mathcal{E}_z$ as large as possible.
Additionally, we need to set the convergence tolerance $E_{\text{tol}}$ of the self-consistent electronic loop (the EDIFF flag in VASP) to a value that is strict enough, to ensure that the resulting dynamical matrix is precise enough to see the effect of the electric field.

The electric field strength $\mathcal{E}_z$ must be chosen well below the onset of Zener breakdown, where the electric enthalpy function loses its minima~\cite{souza2002}. This occurs when:
\begin{equation}
    E^{(\text{Zener})}_z = \frac{\Delta E}{e a N_{\mathbf{k}}} = 0.033~\frac{\text{V}}{\text{\AA}}
\end{equation}
Here, $N_{\mathbf{k}} = 8$ is the number of $\mathbf{k}$-points in the $z$-direction.
To satisfy this criterion, we choose $\mathcal{E}_z = 0.0025~\frac{\text{V}}{\text{\AA}}$ for the calculation of the electric field derivative of the dynamical matrix.

To choose an appropriate value for the convergence tolerance of the self-consistent electronic loop, we introduce a quantitative measure of the numerical error introduced during the first principles calculations.
In~\cite{houtput2025}, it was shown that in the cubic perovskite structure, the dynamical matrices at $+\mathcal{E}_z$ and $-\mathcal{E}_z$ are related by a complex conjugate:
\begin{equation} \label{ComplexConjugateRelation}
\mathcal{D}_{\kappa \alpha, \kappa' \beta}(\mathbf{q}, -\mathcal{E}_z) = \mathcal{D}_{\kappa \alpha, \kappa' \beta}(\mathbf{q}, \mathcal{E}_z)^*.
\end{equation}
From this property, it follows that for a perfect calculation:
\begin{equation}
\text{Im}\left[\mathcal{D}_{\kappa \alpha, \kappa' \beta}(\mathbf{q}_c, \mathcal{E}_z)\right] + \text{Im}\left[\mathcal{D}_{\kappa \alpha, \kappa' \beta}(\mathbf{q}_c, -\mathcal{E}_z)\right] = 0.
\end{equation}
However, in practice, the dynamical matrices at $+\mathcal{E}_z$ and $-\mathcal{E}_z$ are calculated independently, and the above property is not exactly satisfied.
If the result is far from zero, this is an indication of a significant numerical error in the calculation.
We can turn this intuition into a single number by summing over all $\kappa \alpha, \kappa' \beta$, and all commensurate points $\mathbf{q}_c \neq \mathbf{0}$. We exclude the $\Gamma$-point explicitly because $\text{Im}\left[\mathcal{D}_{\kappa \alpha, \kappa' \beta}(\mathbf{0}, \mathcal{E}_z)\right] = 0$.
We introduce the following ``noise metric'' $\chi^2_{\text{imag}}$, which is a dimensionless number:
\begin{widetext}
\begin{align}
\chi^2_{\text{imag}} & := \frac{\underset{\mathbf{q}_c \neq \mathbf{0}}{\sum} \ \underset{\kappa \alpha}{\sum} \ \underset{\kappa' \beta}{\sum} \bigg(\text{Im}\left[\mathcal{D}_{\kappa \alpha, \kappa' \beta}(\mathbf{q}_c, \mathcal{E}_z)\right] + \text{Im}\left[\mathcal{D}_{\kappa \alpha, \kappa' \beta}(\mathbf{q}_c, -\mathcal{E}_z)\right]\bigg)^2}{\underset{\mathbf{q}_c \neq \mathbf{0}}{\text{max}} \ \underset{\kappa \alpha}{\text{max}} \ \underset{\kappa' \beta}{\text{max}} \bigg(\text{Im}\left[\mathcal{D}_{\kappa \alpha, \kappa' \beta}(\mathbf{q}_c, \mathcal{E}_z)\right] - \text{Im}\left[\mathcal{D}_{\kappa \alpha, \kappa' \beta}(\mathbf{q}_c, -\mathcal{E}_z)\right]\bigg)^2 }
\end{align}
We also introduce a similar metric $\chi^2_{\text{der}}$, based on the fact that the derivative of the dynamical matrix should be purely imaginary in the cubic perovskite structure:

\begin{figure*}
    \centering
    \includegraphics[width=16.2cm]{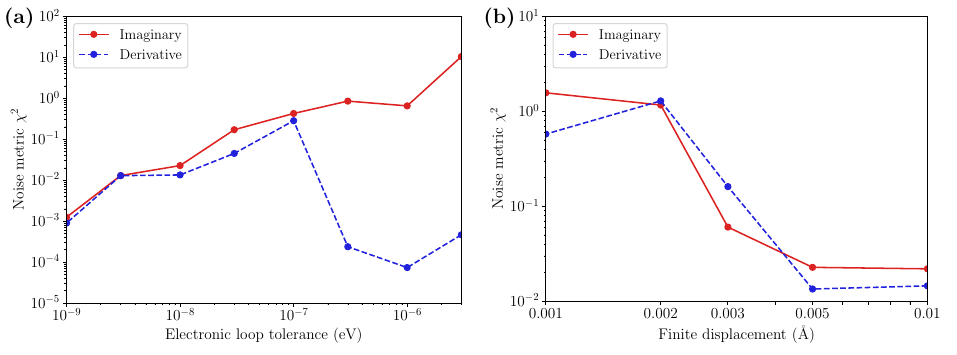}
    \caption{Noise metrics $\chi^2_{\text{imag}}$ and $\chi^2_{\text{der}}$: \textbf{(a)} as a function of the convergence tolerance $E_{\text{tol}}$ of the self-consistent electronic loop, calculated with $\Delta x = 0.005~\text{\AA}$; \textbf{(b)} as a function of the finite displacement $\Delta x$ used to calculate the phonon properties, calculated with $E_{\text{tol}} = 10^{-8}~\text{eV}$. The noise metrics are small enough once $E_{\text{tol}} \lesssim 10^{-8}~\text{eV}$ and $\Delta x \gtrsim 0.005~\text{\AA}$.}
    \label{fig:noise}
\end{figure*}

\begin{align}
\chi^2_{\text{der}} & := \frac{\underset{\mathbf{q}_c \neq \mathbf{0}}{\sum} \ \underset{\kappa \alpha}{\sum} \ \underset{\kappa' \beta}{\sum} \text{Re}\left[\frac{\partial \mathcal{D}_{\kappa \alpha, \kappa' \beta}(\mathbf{q}_c)}{\partial \mathcal{E}_z}\right]^2}{\underset{\mathbf{q}_c \neq \mathbf{0}}{\text{max}} \ \underset{\kappa \alpha}{\text{max}} \ \underset{\kappa' \beta}{\text{max}} \left|\frac{\partial \mathcal{D}_{\kappa \alpha, \kappa' \beta}(\mathbf{q}_c)}{\partial \mathcal{E}_z}\right|^2 }
\end{align}
\end{widetext}
where $\frac{\partial \mathcal{D}_{\kappa \alpha, \kappa' \beta}(\mathbf{q})}{\partial \mathcal{E}_z}$ is calculated using \eref{FiniteDifference}.
Both $\chi^2_{\text{imag}}$ and $\chi^2_{\text{der}}$ can be interpreted as follows.
If $\chi^2_{\text{imag}}$ (resp. $\chi^2_{\text{der}}$) is close to zero, the calculation of the imaginary part (resp. the derivative) of the dynamical matrix is accurate;
the larger $\chi^2$, the more numerical noise is present in the calculations.
If $\chi^2 \approx 1$, we can very roughly estimate that the numerical noise is around the same order of magnitude of the values of the actual elements we are trying to calculate.
This is obviously something to be avoided, so we should aim for a value of $\chi^2 \ll 1$.

In practice, we only use the imaginary part of the dynamical matrix to calculate $\frac{\partial \mathcal{D}_{\kappa \alpha, \kappa' \beta}(\mathbf{q})}{\partial \mathcal{E}_z}$~\cite{houtput2025}.
Indeed, from \eref{ComplexConjugateRelation}, it can be shown that only the imaginary part contributes to \eref{FiniteDifference}.
Therefore, $\chi^2_{\text{imag}}$ is the most important metric of the two.
We use $\chi^2_{\text{der}}$ as a consistency check.
Indeed, the derivative of the dynamical matrix should be equal to the imaginary part of the dynamical matrix, up to a constant prefactor~\cite{houtput2025}.
Therefore, if the calculation is accurate, we also expect $\chi^2_{\text{imag}} \approx \chi^2_{\text{der}}$.

Having introduced the noise metrics $\chi^2_{\text{imag}}$ and $\chi^2_{\text{der}}$, we may now calculate the dynamical matrix at $\mathcal{E}_z = \pm 0.0025~\text{\AA}$ for various values of the ionic displacement magnitude $\Delta x$ and the convergence tolerance $E_{\text{tol}}$ of the self-consistent electronic loop.
\figref{fig:noise} shows the noise metrics $\chi^2_{\text{imag}}$ and $\chi^2_{\text{der}}$ as a function of these two numerical parameters.
For phonon calculations using the VASP + PhonoPy workflow, it is usually recommended to set $E_{\text{tol}} = 10^{-6}~\text{eV}$.
However, from \figref{fig:noise}a it is clear that the results for the electric field calculations will not be reliable with this value, since $\chi^2_{\text{imag}} \approx 1$ and $\chi^2_{\text{imag}} \neq \chi^2_{\text{der}}$.
Instead, we should choose a lower value of $E_{\text{tol}}$, keeping in mind that lower values mean longer computation times.
For all calculations in the main part of the manuscript, we choose the value $E_{\text{tol}} = 10^{-8}~\text{eV}$ as a good compromise.
For clarity, we note that the dynamical matrix $\mathcal{D}_{\kappa \alpha, \kappa' \beta}(\mathbf{q})$ and derived phonon properties will be accurately calculated even at $E_{\text{tol}} = 10^{-6}~\text{eV}$;
it is only for an accurate calculation of the derivative $\frac{\partial \mathcal{D}_{\kappa \alpha, \kappa' \beta}(\mathbf{q})}{\partial \mathcal{E}_z}$ that we need a stricter value.
Similarly, from \figref{fig:noise}b we may conclude that the results for the derivative of the dynamical matrix are reliable for $\Delta x = 0.005~\text{\AA}$, which is the value that we choose for all calculations in this manuscript.
Although we would find even more accurate phonon frequencies by choosing $\Delta x$ even lower, reliable results for $\frac{\partial \mathcal{D}_{\kappa \alpha, \kappa' \beta}(\mathbf{q}_c)}{\partial \mathcal{E}_z}$ would require even lower values of $E_{\text{tol}}$ in this case.

\begin{figure}
\centering
\includegraphics[width=8.6cm]{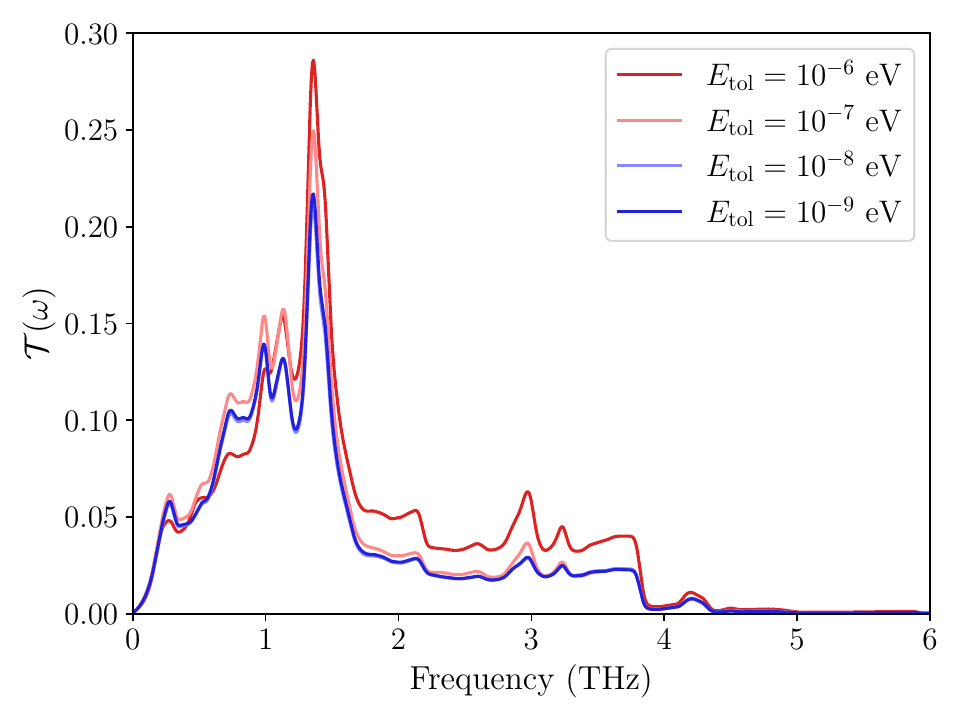}
\caption{\label{fig:Tomega_EDIFF} Convergence of the one-electron-two-phonon spectral function $\mathcal{T}(\omega)$ with respect to the convergence tolerance $E_{\text{tol}}$ of the self-consistent electronic loop.}
\end{figure}

To verify that the physically relevant results are converged with our choice of $E_{\text{tol}}$, we calculated the one-electron-two-phonon spectral function $\mathcal{T}(\omega)$ for several values of this numerical parameter.
\figref{fig:Tomega_EDIFF} shows that $\mathcal{T}(\omega)$ is indeed converged when it is calculated using $E_{\text{tol}} = 10^{-8}~\text{eV}$, since the result is essentially the same as the result calculated using $E_{\text{tol}} = 10^{-9}~\text{eV}$.
It can also be seen that using the usual value $E_{\text{tol}} = 10^{-6}~\text{eV}$ results in a one-electron-two-phonon spectral function that is too large, especially at higher frequencies.

\section{Validity of the long-range approximation} \label{app:LRbenchmarking}

All results in the main part of the manuscript have been obtained using the long-range approximation.
More specifically, this means that we have approximated both the linear and nonlinear matrix elements by only their long-range dipole contributions \cite{houtput2025}:
\begin{align}
g_{mn\nu}(\mathbf{k},\mathbf{q}) & \approx \frac{i e^2}{\epsvac \Omega_0}   \sum_{\mathbf{G} \neq -\mathbf{q}} \sum_{\kappa} \sqrt{\frac{\hbar}{2 m_{\kappa} \omega_{\mathbf{q}\nu}}}  \frac{(\mathbf{q}+\mathbf{G}) \cdot \mathbf{Z}_{\kappa} \cdot \mathbf{e}_{\kappa,\nu}(\mathbf{q}) }{(\mathbf{q}+\mathbf{G}) \cdot \bm{\varepsilon}_{\infty} \cdot (\mathbf{q}+\mathbf{G})} \nonumber \\
& \hspace{170pt} \times \left\langle \psi_{\mathbf{k}+\mathbf{q}m} \left| e^{i (\mathbf{q}+\mathbf{G}) \cdot \mathbf{r}} \right|\psi_{\mathbf{k}n} \right\rangle, \label{g1Long} \\
 g_{mn\nu_1\nu_2}(\mathbf{k},\mathbf{q}_1,\mathbf{q}_2) & \approx \frac{e^2}{2 \epsvac \Omega_0} \sum_{\mathbf{G} \neq -\mathbf{q}_1 - \mathbf{q}_2} \frac{(\mathbf{q}_1+\mathbf{q}_2+\mathbf{G}) \cdot \mathbf{Y}_{\nu_1 \nu_2}(\mathbf{q}_2) }{(\mathbf{q}_1+\mathbf{q}_2+\mathbf{G}) \cdot \bm{\varepsilon}_{\infty} \cdot (\mathbf{q}_1+\mathbf{q}_2+\mathbf{G})}  \nonumber \\
& \hspace{170pt} \times \langle \psi_{\mathbf{k}+\mathbf{q}_1+\mathbf{q}_2m} | e^{i (\mathbf{q}_1+\mathbf{q}_2+\mathbf{G})\cdot \mathbf{r}} | \psi_{\mathbf{k}n} \label{g2Long} \rangle.
\end{align}
Other contributions, i.e. the long-range quadrupole contribution \cite{brunin2020, jhalani2020} and the short-range contribution \cite{giustino2017}, have been neglected.
In practice, this approximation is necessary because there is no theory or code for calculating the quadrupole or short-range part of the nonlinear matrix element as of yet.

A natural question is how accurate this approximation is, compared to using the full electron-phonon matrix elements.
For the nonlinear electron-phonon interaction, such a comparison is not possible since the full $g_{mn\nu_1\nu_2}(\mathbf{k},\mathbf{q}_1,\mathbf{q}_2)$ is not available.
However, the full $g_{mn\nu}(\mathbf{k},\mathbf{q})$ is available in various first-principles codes, so we can verify the accuracy of the long-range approximation on the level of the linear approximation.

Such a benchmark test is performed in \citeref{ponce2019}, where the three LO phonon frequencies $\omega_{\nu}$ and their associated mode polarities $p_{\nu}$ are treated as fitting parameters.
\citeref{ponce2019} shows that the long-range model with these six fitting parameters is an excellent approximation for the mobility calculated with the full electron-phonon matrix element.
However, this is not the method that we use in this manuscript.
In our long-range model, there are no fitting parameters: the required parameters $\omega_{\nu}$, $p_{\nu}$, and $\mathbf{Y}_{\nu_1 \nu_2}(\mathbf{q})$ are all calculated from first principles.
Therefore, in this section, we investigate whether the long-range dipole approximation is still valid when no fitting parameters are used.

\begin{figure}
    \centering
    \includegraphics[width=16.2cm]{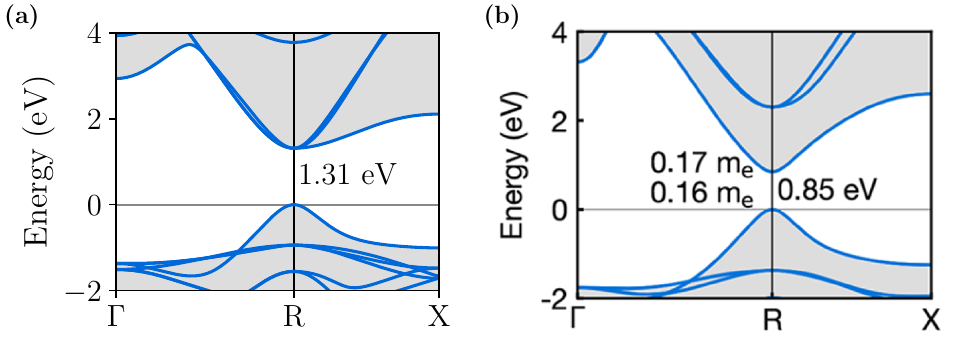}
    \caption{\textbf{(a)} Electron bands $\varepsilon_{\mathbf{k}n}$ calculated with the DFT parameters of this manuscript, as described in \secref{appsec:DFTelectronic}. Notably, this calculation ignores spin-orbit coupling. \textbf{(b)} Figure reproduced from \citeref{ponce2019}, which shows the same electron bands calculated with spin-orbit coupling and a GW correction.}
    \label{fig:CsPbI3_elbands}
\end{figure}

For this benchmark, we will only compare the hole mobility of CsPbI$_3$.
This is because we perform all of our calculations without spin-orbit coupling, which means we find the electron bands shown in \figref{fig:CsPbI3_elbands}a.
In this case, the bottom of the conduction band is threefold degenerate, which invalidates the single band long-range model derived in \secref{app:Mobility}.
In practice this degeneracy is lifted by a remarkably large spin-orbit splitting \cite{ponce2019}, as shown in \figref{fig:CsPbI3_elbands}b.
This is why we are able to use the single band model with $m^*_e = 0.17 m_{\text{el}}$ and $m^*_h = 0.16 m_{\text{el}}$ in the main text.
However, for this benchmark we must calculate everything at the same level of approximation.
Fortunately, the top of the valence band still consists of a single parabolic band, with a band mass we calculated to be $m^*_h = 0.124 m_{\text{el}}$.
This is why we will only calculate the hole mobility of CsPbI$_3$ for this benchmark, so we can use both the long-range model as well as the full first-principles calculation.
As also explained in the main text, we expect this benchmark to be representative for the carrier mobility as well, since the electron band shape only enters the mobility through the band mass, as a constant prefactor $\mu \sim (m^*)^{-\frac{3}{2}}$ (see \eref{muSERTApractical}).

We use VASP v.6.5.1 ~\cite{kresse1996, kresse1996a} to calculate the full electron-phonon matrix element $g_{mn\nu}(\mathbf{k},\mathbf{q})$ and the derived SERTA mobility.
The calculation comprises two steps. First, there is a supercell calculation to calculate the phonon properties and the electron-phonon potential: for this calculation, we use the same parameters as those defined in \secref{app:DFTinputs}.
Then, there is a unit cell calculation where the supercell electron-phonon potential is used to evaluate the $g_{mn\nu}(\mathbf{k},\mathbf{q})$, the imaginary part of the Fan-Migdal self energy (equivalent to the electron lifetimes), and finally the SERTA mobility.
The unit cell calculation consists of a self-consistent calculation on an $8 \times 8 \times 8$ $\mathbf{k}$-grid, followed by a one-shot calculation where the electron bands $\varepsilon_{\mathbf{k}n}$, phonon frequencies $\omega_{\mathbf{q}\nu}$, and electron-phonon matrix elements $g_{mn\nu}(\mathbf{k},\mathbf{q})$ are calculated on a dense $N_{\mathbf{q}} \times N_{\mathbf{q}} \times N_{\mathbf{q}}$ $\mathbf{k}$- and $\mathbf{q}$-grid.
Any $\mathbf{q}$-points that include imaginary phonon modes (see \figref{fig:Inputs}) are ignored in the calculation: this is an excellent approximation since we only consider the linear electron-phonon interaction in this section, which only requires phonons with $\mathbf{q}$ close to $\Gamma$.
In order to calculate the intrinsic mobility, we choose to perform the calculation at a low hole doping density of $n_h = 10^{15}~\text{cm}^{-3}$.

The $\mathbf{q}$-grid density $N_{\mathbf{q}}$ must be chosen large enough such that convergence is reached.
Before starting calculations, it is useful to estimate the value of $N_{\mathbf{q}}$ that is required for convergence.
The SERTA mobility with the full electron-phonon matrix element is given by \erefs{muSERTA}{ooeoF}, where the electron lifetimes are calculated as \cite{ponce2020}:
\begin{multline}
\frac{1}{\tau_{\mathbf{k}n}} = \frac{2\pi}{\hbar} \sum_{m \nu} \frac{\Omega_0}{(2\pi)^3} \int_{\text{1BZ}}  \dee^3 \mathbf{q} \ |g_{mn\nu}(\mathbf{k},\mathbf{q})|^2 \times \\
\bigg\{
\big[1 - f^0(\varepsilon_{\mathbf{k}+\mathbf{q}m}) + n(\hbar \omega_{\mathbf{q}\nu})\big] \delta(\varepsilon_{\mathbf{k}n} - \varepsilon_{\mathbf{k}+\mathbf{q}m} - \hbar \omega_{\mathbf{q}\nu}) \\
 + \big[f^0(\varepsilon_{\mathbf{k}+\mathbf{q}m}) + n(\hbar \omega_{\mathbf{q}\nu})\big]\delta(\varepsilon_{\mathbf{k}n} - \varepsilon_{\mathbf{k}+\mathbf{q}m} + \hbar \omega_{\mathbf{q}\nu})
  \bigg\} .
\end{multline}
For the hole mobility, contributions to $\tau_{\mathbf{k}n}$ must come from electron-phonon processes that satisfy the last Dirac delta function:
\begin{equation}
\varepsilon_{\mathbf{k}n} - \varepsilon_{\mathbf{k}+\mathbf{q}m} + \hbar \omega_{\mathbf{q}\nu} = 0
\end{equation}
The crucial insight is that the phonon frequencies in CsPbI$_3$ are much smaller than the electron energies, $\hbar \omega_{\mathbf{q}\nu} \lesssim 16~\text{meV}$.
Therefore, we only get contributions within the same band ($n=m$), and if the $n_h$-doping density is low, this band must be the valence band maximum.
If we plug in the parabolic expression for the valance band, we find that $\mathbf{q}$ contributes to the electron lifetime when:
\begin{equation}
\frac{\hbar^2 |\mathbf{q}|^2}{2m^*_h} = \hbar \omega_{\mathbf{q}\nu}
\end{equation}
Since the solution of this equation satisfies $\mathbf{q} \approx \mathbf{0}$, we may evaluate the phonon frequencies $\omega_{\nu}$ at $\Gamma$.
Then, we expect that a given $\mathbf{q}$-point can contribute to $\tau_{\mathbf{k}n}$ if:
\begin{equation} \label{qArgument}
 \Leftrightarrow |\mathbf{q}| \frac{a}{2\pi} = \frac{a}{2\pi} \sqrt{\frac{2m^*_h \omega_{\nu}}{\hbar}} \approx \left\{
\begin{array}{lcl}
\displaystyle \frac{1}{120} & \text{ for } & \omega_{\nu} = \omega_{\text{LO}1}, \\[7pt]
\displaystyle \frac{1}{90} & \text{ for } & \omega_{\nu} = \omega_{\text{LO}2}, \\[7pt]
\displaystyle \frac{1}{45} & \text{ for } & \omega_{\nu} = \omega_{\text{LO}3}.
\end{array} 
 \right.
\end{equation}
In other words, we expect to require at least $N_{\mathbf{q}} \gtrsim 45$ before we start to properly see the contribution from the highest LO phonon frequency, and we require a very dense grid with at least $N_{\mathbf{q}} \gtrsim 120$ to properly see all required contributions.
In other materials with higher phonon frequencies $\omega_{\nu}$ or a higher band mass $m^*$, the required $N_{\mathbf{q}}$ would be lower.
This reasoning is also a strong argument in favor of the validity of the long-range approximation in CsPbI$_3$, since it shows that only phonons with $|\mathbf{q}|a \ll 1$ can contribute to the mobility.
We highlight that in the long-range model, the integral over $\mathbf{q}$ is calculated analytically (see \eref{arcsinIntegral}).
This is a big computational advantage over the calculation with the full $g_{mn\nu}(\mathbf{k},\mathbf{q})$, which requires a very dense $N_{\mathbf{q}} \times N_{\mathbf{q}} \times N_{\mathbf{q}}$ grid.

Since the $\mathbf{q}$-grid needs to be very dense, we choose the remaining numerical parameters to be less strict than their default values, in order to save computation time and memory which can go into increasing $N_{\mathbf{q}}$.
The plane-wave cutoff energy is set to $250~\text{eV}$, which saves memory compared to using the higher value of $600~\text{eV}$.
We include the long-range contribution in the calculation of the electron-phonon matrix element;
to save computation time, it turned out to be particularly important to set the reciprocal space sum cutoff in this long-range contribution to $8~\text{eV}$, instead of the default value of $50~\text{eV}$.
Finally, the number of points in the Gauss-Legendre grid for evaluating energy integrals over the Fermi-Dirac distribution and the transport distribution function were both left at the default value of 501, since these could not be lowered without meaningfully changing the resulting mobility.
We have verified that choosing these values gave mobility values within a few percent of those calculated with the default numerical parameters.

\begin{figure}
    \centering
    \includegraphics[width=16.2cm]{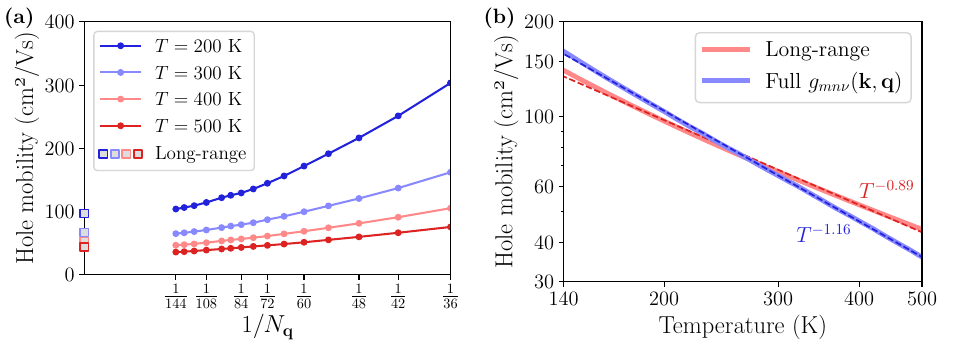}
    \caption{Hole mobility of CsPbI$_3$ in the linear electron-phonon approximation, calculated using both fully from first principles and using the long-range approximation from the main text.
    \textbf{(a)} Convergence with respect to the density of the $\mathbf{k}$-grid and $\mathbf{q}$-grid, $N_{\mathbf{q}} \times N_{\mathbf{q}} \times N_{\mathbf{q}}$.
    \textbf{(b)} Comparison of the temperature dependence between the first principles calculations and the long-range approximation, calculated on a $144 \times 144 \times 144$ $\mathbf{k}$-grid and $\mathbf{q}$-grid.
    Dashed lines indicate power law fits of the form $\mu \sim T^{-\gamma}$.}
    \label{fig:CsPbI3_LRbenchmark}
\end{figure}

With the numerical parameters fully defined as above, we calculate hole mobility as a function of temperature and $N_{\mathbf{q}}$.
\figref{fig:CsPbI3_LRbenchmark}a shows the hole mobility for values up to $N_{\mathbf{q}} = 144$.
At this value, the hole mobility is not varying too much anymore, indicating that the result is converged.
This is in accordance with our estimate that we require $N_{\mathbf{q}} \gtrsim 120$ for convergence.
Therefore, we choose a $144 \times 144 \times 144$ $\mathbf{q}$-grid for our final analysis.

\figref{fig:CsPbI3_LRbenchmark}b shows the comparison between first-principles mobility calculation and the mobility in the long-range model, both as a function of temperature.
It is clear that the long-range approximation matches qualitatively with the first-principles calculation.
In particular, the absolute values of the mobility are not too different, indicating that it is safe to use the long-range approximation to investigate the relative effect of the nonlinear electron-phonon interaction.
However, it is also important to note that the temperature scaling of the long-range model is not quite correct.
Both calculations predict a power law dependence on temperature of the form $\mu \sim T^{-\gamma}$, but the long-range model predicts $\gamma = 0.89$ and the first-principles calculation predicts $\gamma = 1.16$.
This discrepancy is likely due to a long-range quadrupole contribution or a short-range contribution, which is included in the first-principles calculation but not in the long-range dipole approximation.
From the argument preceding \eref{qArgument}, it seems likely that the long-range quadrupole contribution around $\mathbf{q}=\mathbf{0}$ provides most of the missing contribution, though a possible short-range contribution cannot be excluded.

We conclude this section by connecting the results of \figref{fig:CsPbI3_LRbenchmark} to the nonlinear electron-phonon interaction in the main part of the manuscript.
Firstly, we note that as long as there is no theory for the quadrupole or short-range contributions to the one-electron-two-phonon matrix element, any first-principles investigation of the halide perovskites has to choose between making the long-range dipole approximation or the linear approximation.
Comparing the results of \figref{fig:Mobility} and \figref{fig:CsPbI3_LRbenchmark}, we see that including the full linear electron-phonon matrix element ($\gamma = 1.16$) is more impactful than including long-range one-electron-two-phonon interaction ($\gamma = 0.97$).
Therefore, if a choice must be made between these two approximations, it makes sense to use the linear electron-phonon coupling theory that is established in the literature.
Secondly, however, we emphasize that both including nonlinear electron-phonon interaction and using the full electron-phonon matrix element must be included in principle.
In fact, both will bring the power law scaling $\gamma$ closer to the experimental value in single-crystal MAPbI$_3$ from \figref{fig:Mobility} \cite{xia2021}.
Hypothetically, a theory that includes the full linear and nonlinear electron-phonon matrix elements might provide a temperature dependence that matches experiments.
Finally, the most important conclusion from the main part of this manuscript is that the nonlinear electron-phonon interaction decreases the phonon-limited mobility by about $10\%$ at room temperature.
Since \figref{fig:CsPbI3_LRbenchmark}b shows that the room-temperature mobility is correctly predicted by the long-range dipole approximation, this important conclusion remains valid even when short-range interaction is included.

\bibliography{References}

% We manually include main.aux so the cross-referencing works